\documentclass{article}

\usepackage{graphicx} 
\usepackage{booktabs}
\usepackage{float}
\usepackage{setspace}
\usepackage{adjustbox}
\usepackage{threeparttable}
\usepackage{multirow, makecell}
\usepackage{PRIMEarxiv}
\usepackage{wrapfig,lipsum,booktabs}

\usepackage[utf8]{inputenc} 
\usepackage[T1]{fontenc}    
\usepackage{hyperref}       
\usepackage{url}            
\usepackage{booktabs}       
\usepackage{amsfonts}       
\usepackage{amsmath} 
\usepackage{multirow}
\usepackage{array}
\usepackage{nicefrac}       
\usepackage{microtype}      
\usepackage{lipsum}
\usepackage{tcolorbox}
\usepackage{lmodern}
\usepackage{fancyhdr}       
\usepackage{graphicx}       
\usepackage{listings}       
\usepackage{chngcntr}       
\graphicspath{{media/}}     

\newcommand{\llamattwo}{\textsc{LLaMat-2}{ }}
\newcommand{\llamatthree}{\textsc{LLaMat-3}{ }}
\newcommand{\llamatwo}{\textsc{LLaMA-2}{}}
\newcommand{\llamathree}{\textsc{LLaMA-3}{}}
\newcommand{\llama}{\textsc{LLaMA}{}}
\newcommand{\llamat}{\textsc{LLaMat}{}}

\title{Foundational Large Language Models for Materials Research
}

\author{
    Vaibhav Mishra$^{1,*}$, Somaditya Singh$^{1,*}$, Dhruv Ahlawat$^{1,*}$, Mohd Zaki$^{2,*}$, \\ 
    \textbf{Vaibhav Bihani}$^{3}$, \textbf{Hargun Singh Grover}$^{3}$, \textbf{Biswajit Mishra}$^{4}$, \textbf{Santiago Miret}$^{5}$,\\
    \textbf{Mausam}$^{1,3,\#}$, \textbf{N. M. Anoop Krishnan}$^{2,3,\#}$ \\
    $^{1}$Department of Computer Science and Engineering, $^{2}$Department of Civil Engineering \\
    $^{3}$Yardi School of Artificial Intelligence, Indian Institute of Technology Delhi \\
    $^{4}$Cerebras Systems, Inc., 
    $^{5}$Intel labs \\
    $^\#$Corresponding authors: \texttt{\{mausam, krishnan\}@iitd.ac.in}\\
    $^*$Authors contributed equally.} 

\tcbuselibrary{breakable}
\begin{document}
\maketitle

\begin{abstract}
Materials discovery and development are critical for addressing global challenges in renewable energy, sustainability, and advanced technology. Yet, the exponential growth in materials science literature comprising vast amounts of textual data has created significant bottlenecks in knowledge extraction, synthesis, and scientific reasoning. Large Language Models (LLMs) offer unprecedented opportunities to accelerate materials research through automated analysis and prediction. Still, their effective deployment for materials discovery requires domain-specific adaptation for language understanding and solving domain-relevant tasks. Here, we present \llamat{}, a family of foundational models for materials science, developed through continued pretraining of \llama{} models on an extensive corpus of materials literature and crystallographic data, followed by instruction- and task-finetuning. Through systematic evaluation, we demonstrate that \llamat{} excels in materials-specific natural language processing and structured information extraction tasks outperforming commercial LLMs, while maintaining general linguistic capabilities. The specialized \llamat{-CIF} variant demonstrates remarkable capabilities in crystal structure generation, predicting stable crystals with high coverage across the periodic table. Intriguingly, despite \llamathree{'s} superior performance in comparison to \llamatwo{}, we observe that \llamattwo{} demonstrates unexpectedly enhanced domain-specific performance across diverse materials science tasks, including structured information extraction from text and tables and crystal structure generation. These results point to a potential ``adaptation rigidity'' in overtrained LLMs such as \llamathree{}. Altogether, the present work demonstrates the effectiveness of domain adaptation towards the development of practically deployable LLM copilots for materials research. Beyond materials science, our findings reveal important considerations for domain adaptation of LLMs---model selection, training methodology, and domain-specific performance---that may influence the development of specialized scientific AI systems.
\end{abstract}


\section{Introduction}
Materials innovation can potentially address ten of the seventeen United Nations Sustainable Development Goals through advances in sustainable energy systems, advanced electronics, and environmentally conscious manufacturing. This imperative for accelerated materials discovery coincides with an unprecedented expansion in the scientific literature---exceeding 6 million materials science (MatSci) publications---presenting both opportunities and challenges for materials informatics~\cite{krishnan2024machine,venugopal_matkg_2024-1,miret2024llms}. Obtaining actionable insights from this big-data requires advanced computational tools that can effectively process vast scientific literature, the majority of which is unstructured text data and semi-structured tables.

Large language models (LLMs), also referred to as foundation models, have demonstrated remarkable capabilities in text processing, analysis, and generation~\cite{bubeck2023sparks}. In the field of materials, LLMs can enhance the research and discovery process through (i) rapid literature-based identification of materials~\cite{gupta2022matscibert,schilling2024text} and synthesis pathways~\cite{mysore2019materials}, (ii) \emph{in silico} crystal structure generation~\cite{antunes_crystal_2024, gruver2024fine, ding2024matexpert}, (iii) autonomous experimental planning~\cite{autolab,chemcrow, sim2024chemos}, and (iv) results analysis~\cite{zaki_mascqa, chemistry_white2023assessment}. Recent advances~\cite{dagdelen_structured_2024, sayeed2024nlp-taylor,alampara2024mattext,hackathon-2024-zimmermann2024reflections} have demonstrated the efficacy of LLMs in materials concept comprehension, domain-specific query resolution~\cite{alampara2024mattext,mirza2024large, zhang2024honeycomb}, and simulation code generation~\cite{zaki_mascqa}. However, critical analyses of the performance of these general-purpose LLMs reveal their inability to address domain-specific challenges, including the incorrect interpretation of scientific phenomena such as physical laws or theories, specialized terminologies~\cite{miret2024llms,alampara2024mattext,hira,zaki_mascqa}, and crystal structures~\cite{antunes_crystal_2024,alampara2024probing}.

Effectively leveraging LLMs for materials research requires specialized domain adaptation to address their limitations in materials-specific information processing~\cite{miret2024llms}. Initial efforts toward domain adaptation of LLMs by fine-tuning them for specific tasks in materials research have yielded promising breakthroughs in structured information extraction~\cite{nerre}, materials-specific natural language processing~\cite{song2023honeybee,dagdelen_structured_2024,sayeed_annotating_2024}, experimental data analysis~\cite{circi2024well,hira}, and crystal structure generation~\cite{gruver2024fine,antunes_crystal_2024,baird2024matbench,ding2024matexpert}. These achievements highlight the potential for a unified materials foundation model that integrates these capabilities to accelerate research and development.

Here, we introduce \llamat{}---a family of domain-adapted language models demonstrating generalist material science capabilities. Through a systematic approach combining pretraining, instruction fine-tuning, and task-specific fine-tuning, \llamat{} enables advanced scientific natural language processing, information extraction, and crystal generation. Our comprehensive evaluation demonstrates that \llamat{} outperform existing LLMs across diverse MatSci tasks and exhibit capabilities that bridge the gap between human expertise and automated materials discovery.
\section{Results}
\subsection{\llamat{}: A Family of Large Language Model for Materials}
\begin{figure}
    \centering
    \includegraphics[width=0.8\linewidth]{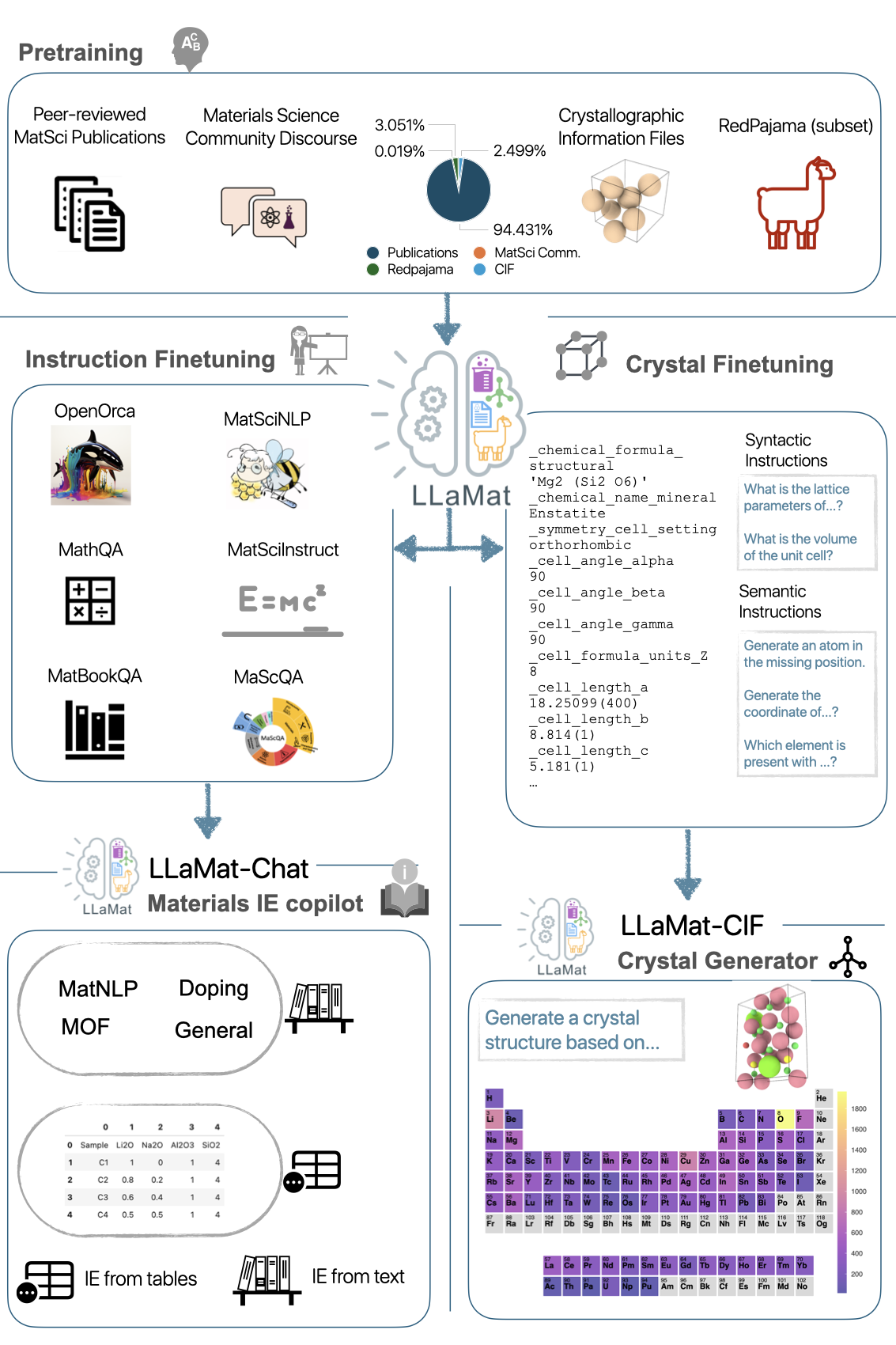}
    \caption{\textbf{Development pipeline and capabilities of \llamat{} for MatSci applications.} The schematic illustrates the two-stage development of \llamat, beginning with continuous pretraining on MatSci corpora (top), followed by specialized instruction finetuning pathways (left and right). The pretraining dataset composition is shown in the pie chart, comprising peer-reviewed publications (94.43\%), crystallographic information files (CIF, 2.50\%), and a subset of RedPajama (3.051\%). Two distinct finetuning pathways yield \llamat{-Chat}, a materials research copilot capable of structured information extraction and materials NLP tasks (left branch), and \llamat{-CIF}, specialized in crystal structure analysis and generation (right branch). Representative examples demonstrate the dataset details and model's capabilities in handling diverse MatSci queries and tasks.}
    \label{fig:llamat}
\end{figure}

To develop \llamat{}, we systematically embedded materials domain knowledge on \llama{} base models---specifically \llamatwo{-7B} \cite{llama2} and \llamathree{-8B}\cite{llama3}, hereafter referred to as \llamatwo{} and \llamathree, respectively. While larger \llama{} variants such as the 70B models could yield superior performance, our model selection optimizes the balance between computational demands for training and inference, available pretraining data volume, and practical deployment considerations for the larger materials community. \llamat{} is developed employing a rigorously designed three-stage pretraining-finetuning process (see Fig.~\ref{fig:llamat}, Methods). The initial stage comprised continued pretraining (CPT) on the base \llama{} models with an extensive and meticulously curated corpus developed in-house, namely R2CID (see Methods and Tab.~\ref{appendix:ift_pretrain_datasets} in App.~\ref{app:dataset} for details) with greater than 30 billion tokens of MatSci knowledge, encompassing approximately 4 million peer-reviewed publications (94.43\%), crystallographic information files (2.499\%), and MatSci community discourse (0.019\%). Additionally, we incorporated a strategic 3\% subset of \textsc{RedPajama} data, the original training corpus of \llama{} models, to preserve fundamental linguistic capabilities while concurrently mitigating catastrophic forgetting. 

Subsequently, we implemented two distinct finetuning pathways to develop specialized \llamat{} variants. The first variant, \llamat-Chat, underwent comprehensive instruction finetuning (IFT) across multiple domains, including general English comprehension, mathematical reasoning, and MatSci-specific datasets (see Methods and App.~\ref{appendix:ift_pretrain_datasets}). This model was further finetuned on a single corpus comprising several materials-relevant downstream tasks  (see Tab.~\ref{appendix:task_descriptions}), resulting in a materials research copilot with demonstrated proficiency in natural language tasks related to MatSci, including named entity recognition, relation classification, and text classification to name a few, as well as structured information extraction from scientific text and tables (App.~\ref{app:downstream_data}). Concurrently, we developed \llamat-CIF models through IFT of \llamat{} models on crystallographic information files, a hand-curated dataset comprising five syntactic and four semantic tasks. Following this, parameter-efficient finetuning (PEFT) was employed on \llamat{-CIF} to enable crystal generation, a task of importance in materials discovery (see Methods Sec.\ref{sec:Methodology} for details).

To obtain the best-performing models, we conducted extensive experiments balancing the datasets, both during CPT and IFT (Appendix~\ref{appendix:section:dataset_optimization}) with the goal of developing a model that performs best on MatSci tasks without losing its original English capabilities. In CPT, we explored several dataset combinations by prioritizing papers and interspersing it with RedPajama and CIF (see Methods). In IFT, we included datasets on general English comprehension (using OpenOrca\cite{mukherjee2023orca}) and mathematical reasoning (using MathQA\cite{math}) alongside MatSci-specific tasks (see App.~\ref{appendix:section:dataset_optimization}), including datasets from MatSciNLP~\cite{song-etal-2023-matsci} and in-house hand-curated datasets on question-answering related to materials domains including MatBookQA (3000 QA pairs), MaScQA (2000 QA pairs), and MatSciInstruct (170k QA pairs)~\cite{song-etal-2023-honeybee} (see App.~\ref{app:dataset}). 

Systematic evaluation of model performance during CPT and IFT stages revealed several notable insights into the domain adaptation process of LLMs. Dataset distribution and learning rates (see App.~\ref{app:hyperparameters}) were found to play a crucial role in governing the model performance. More importantly, the hyperparameters and dataset distribution influencing model performance were found to be distinct for \llamatwo{} and \llamathree{} (see Apps.~\ref{app:hyperparameters} and \ref{appendix:section:dataset_optimization}). Compared to intermediate checkpoints, CPT on the complete domain-specific corpus consistently demonstrated superior performance metrics for both \llamatwo{} and \llamathree{} architectures. During IFT on OpenOrca, model-specific behavioral patterns were observed: while \llamatwo{} showed substantial improvements across evaluation metrics, \llamathree{} demonstrated minimal performance gains across MatSci and general language tasks (see Tab.~\ref{tab:orca_llamat2}). Models trained without MathQA~\cite{math} in their finetuning regime exhibited severe degradation in mathematical reasoning capabilities—failing to solve even elementary arithmetic problems despite maintaining reasonable linguistic performance relative to their respective base models (Tab.~\ref{tab:hbmx}). This finding underscores the importance of having datasets pertaining to diverse capabilities during the domain adaptation process.

Interestingly, following the IFT on OpenOrca and MathQA, additional IFT of \llamat{} models on materials-specific datasets, such as Honeybee~\cite{song-etal-2023-honeybee}, did not yield significant performance improvements of \llamat{} models on either English or MatSci tasks (see Tab.~\ref{tab:hbmx} in Appendix). Note that finetuning of \llamatwo{} on Honeybee had demonstrated significant performance improvements in an earlier study~\cite{song-etal-2023-honeybee}. This unexpected observation suggests a fundamental distinction between domain knowledge acquisition and instruction-following capabilities: while domain adaptation through pretraining and finetuning effectively enhances field-specific performance, the development of robust instruction-following competency appears to be independently trainable through generic question-answer datasets. Through rigorous parametric optimization studies, we identified Pareto-optimal dataset configurations for each base model, effectively maximizing MatSci task performance while maintaining robust general language capabilities (App. ~\ref{appendix:section:dataset_optimization}).

\subsection{Materials Research Copilot}
To assess the model's efficacy as a materials research copilot, we conducted systematic evaluations across two critical domains: Materials' Natural Language Processing (MatNLP) and Materials' Structured Information Extraction (MatSIE). These evaluations are specifically targeted to evaluate the model's ability to comprehend complex MatSci concepts and extract structured information from both textual and tabular data in scientific publications, representing fundamental capabilities required for materials research automation.

\paragraph{Materials Language Processing.} MatNLP encompasses fourteen tasks across three fundamental natural language processing task families: entity recognition, extraction, and classification. The evaluation framework comprises ten materials-specific and four English datasets, totaling 14,579 test instances. These tasks systematically assess the model's capability to extract granular information from materials literature—including synthesis protocols, characterization methods, and application-specific entities. They also include classification tasks (for instance, whether a particular document is related to a topic in materials) and entity relationship comprehension. The English dataset provides a complementary assessment of general language capabilities through question-answering and multiple-choice tasks.

\begin{figure}
    \centering
    \includegraphics[width=0.8\linewidth]{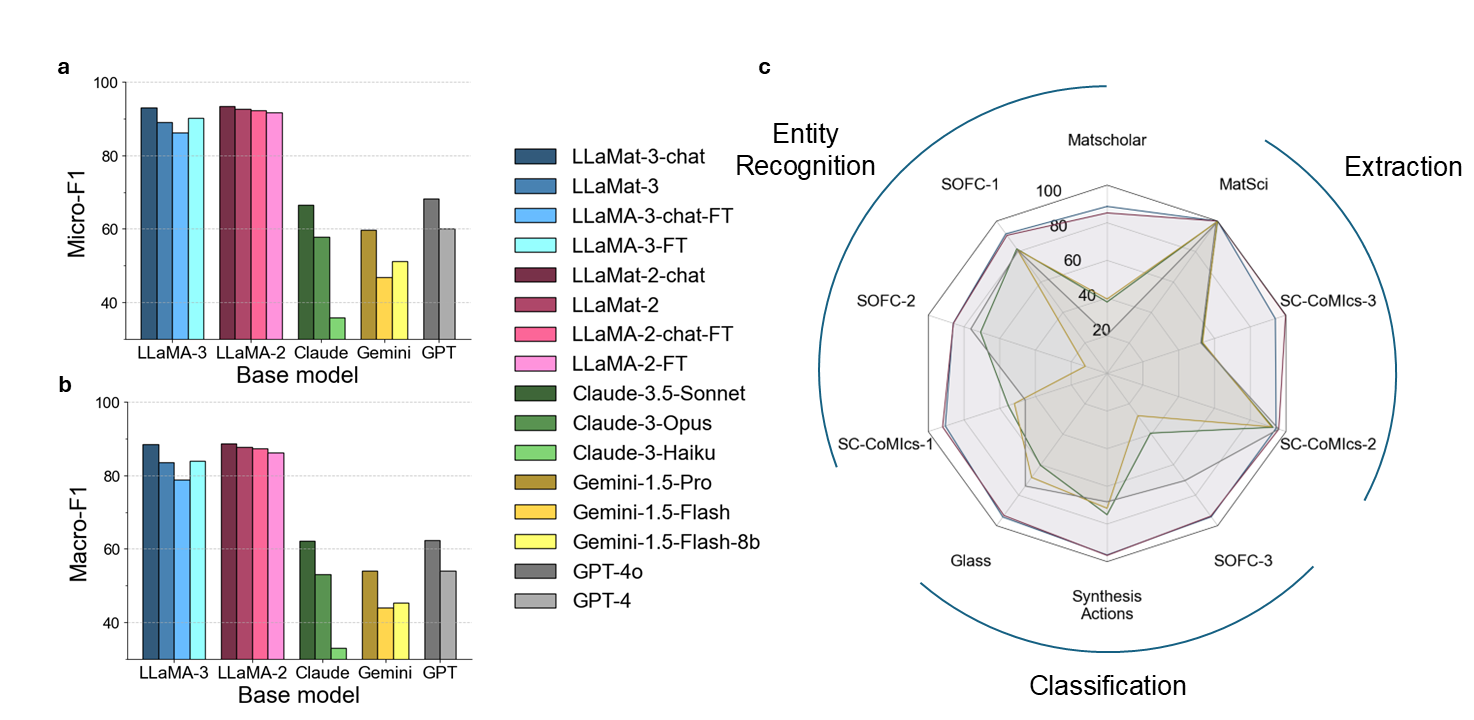}
    \caption{\textbf{Comparative performance analysis of LLaMat and LLaMA models across MatSci and general language tasks with closed source models: Claude and Gemini. LLaMA-FT models correspond to the meta-LLaMA models finetuned on our training corpus} a, Micro-F1, and b, Macro-F1 scores demonstrate performance on MatSci tasks. c, Radar plot illustrating task-specific performance across diverse MatSci applications, including entity recognition, relation extraction, and classification tasks. Only the top models from each family are included in the radar plot, \llamat{-3-chat}, \llamat{-2-chat}, Claude-3.5-Sonnet, Gemini-1.5-Pro, and GPT-4o. For MatSci tasks, higher scores indicate better performance in extracting domain-specific information, identifying relationships between materials entities, and classifying scientific text. Results demonstrate that domain-specific pretraining enhances MatSci task performance while preserving general language capabilities.}
    \label{fig:performance_downstream}
\end{figure}

We evaluate the performance of \llamat{-2} and -3 models and their chat variants on this dataset and compare them to their respective base models and variants of several widely used closed-source models namely, GPT, Claude, and Gemini. For a fair comparison, we also finetuned (FT) both pretrained and chat (instruct) variants of \llamatwo{} and \llamathree{} on the training dataset of downstream task. Figure \ref{fig:performance_downstream} presents a comprehensive performance analysis of \llamat{} in comparison to the finetuned \llama{} variants. The micro and macro F1 scores (Figures~\ref{fig:performance_downstream}a,b) reveal \llamatthree{-Chat}'s superior performance compared to non-chat variants, demonstrating the effectiveness of our domain-specific CPT-IFT strategy. Further, the performance of closed source models is significantly inferior to \llamat{} (see App.~\ref{appendix:section:all_results}). Inference for all the models were performed with a temperature setting of 0.

Our performance analysis reveals interesting architectural dependencies in domain adaptation capabilities. While \llamatthree{} variants show greater relative improvement from their base model compared to \llamattwo{} implementations, the finetuned \llamattwo{} models consistently outperform their \llamathree{} counterparts. This counterintuitive pattern persists even in CPT models without IFT, where \llamattwo{} demonstrates superior performance. This observation suggests a potential domain adaptation limitation in \llamathree{}, possibly stemming from its extensive pretraining ($\sim$3 orders of magnitude more data) despite superior base model performance. This phenomenon, referred to hereafter as ``adaptation rigidity,'' a recurring observation as discussed in later results, underscores the complex relationship between model architecture, pretraining scale, and domain adaptation efficacy\cite{rigid_1, rigid_2}. 


The radar plot (Figure \ref{fig:performance_downstream}e) provides a granular analysis of micro-F1 scores across MatNLP dataset subsets. For both closed-source and our models, only the best among each family is considered. Most notably, \llamatthree{-Chat} model demonstrates consistent performance advantages across diverse MatSci tasks, including entity recognition, classification, and extraction tasks, with \llamattwo{-Chat} is ranked second. All the state-of-the-art commercial models exhibit significantly inferior performance establishing the efficacy of \llamat{} models for broader MatSci applications.

\paragraph{Structured Information Extraction from Text.} The MatSci literature contains vast amounts of information about material compositions, synthesis protocols, and properties embedded within unstructured text. Extracting this information in a structured format is an important step for accelerating materials discovery. Conventionally, this step requires extensive manual annotation and specialized model development for each extraction task. The challenge is particularly acute in specialized domains such as doping studies and metal-organic frameworks (MOFs), where precise extraction of chemical compositions, structural relationships, and functional properties is crucial. While recent studies have demonstrated the potential of finetuned commercial LLMs for these tasks~\cite{dagdelen_structured_2024,morgan}, their proprietary nature and associated costs limit scalable deployment across the millions of articles in materials literature, necessitating the development of open-source alternatives optimized for MatSci applications~\cite{schilling2024text}.

\begin{figure}
    \centering
    \includegraphics[width=0.8\linewidth]{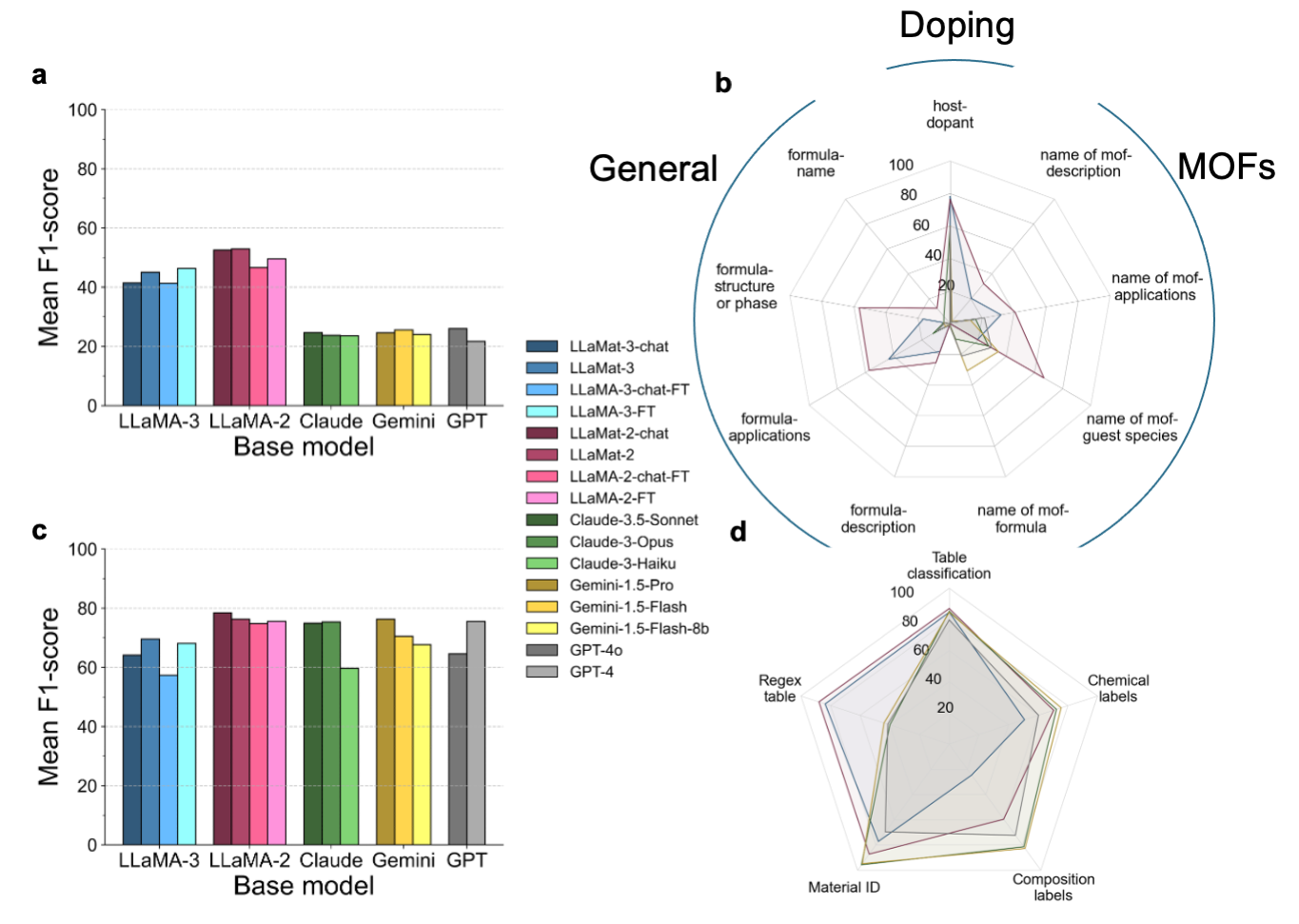}
    \caption{\textbf{Performance evaluation of structured information extraction capabilities across MatSci subdomains.} a, Bar plot showing mean 
    F1 score across all our structured information extraction tasks in doping, metal-organic-frameworks, and general material science, b, Radar plot for F1-score across all relation extraction tasks c, Bar plot showing mean accuracy over all material science table data extraction tasks, d, Radar plot showing F1-score for individual tasks in table data extraction.  Only the top models from each family are included in the radar plots, \llamat{-3-chat}, \llamat{-2-chat}, Claude-3.5-Sonnet, Gemini-1.5-Pro, and GPT-4o}
    \label{fig:performance_matsie}
\end{figure}

Having established the superior performance of \llamat{-Chat} models in MatNLP tasks, we now evaluated their structured information extraction capabilities. Figure \ref{fig:performance_matsie}a demonstrates the performance of \llamat{-Chat} models and closed source models across nine distinct extraction tasks in the doping, metal-organic framework (MOF), and general materials domains showcasing better capabilities of the former compared to closed source models. We observe that \llamat{} models again outperform all the state-of-the-art LLMs. The results of all the variants of \llama{} and \llamat{} models along with the LLMs are provided in App.~\ref{appendix:section:all_results}. Further, both \llamat{-2} and \llamat{-3} chat variants consistently outperform their finetuned \llama{} counterparts in extracting relationships between host materials and dopants, formula-structure mappings, and application-specific information. \ref{fig:performance_matsie}b shows the performance of the best-in-class models for each family. We note that \llamat{-Chat} models consistently outperform all other models across most among the nine MatSIE tasks. 

Notably, \llamat{-2}-chat exhibits particularly strong performance in formula-application relationships and host-dopant associations. This performance pattern aligns with our earlier observations of the adaptation rigidity phenomenon. Specifically, \llamattwo{-Chat} model exhibits significantly enhanced capabilities compared to its successor after domain adaptation through CPT and IFT. This consistent trend across evaluation metrics reinforces our hypothesis about the inverse relationship between the initial pretraining scale and domain adaptation efficacy.

\paragraph{Information Extraction from Tables.}
Tables in the materials domain serve as structured repositories of composition--property data yet present unique challenges due to their heterogeneous formats and complex organizational schemas across publications~\cite{gupta-etal-2023-discomat, hira}. This inherent variability in tabular data representation demands advanced language models capable of understanding the context of MatSci and extracting structured information with high fidelity.

We now evaluate the capability of \llamat{} models to extract meaningful information from materials tables. To this end, we consider five critical capabilities: compositional table classification, chemical constituent localization, composition extraction, material identifier recognition, and regex-amenable information identification. We consider a set of 737 tables from peer-reviewed publications to evaluate the same. These tables were presented in a challenging manually annotated benchmark dataset for information extraction from tables~\cite{gupta-etal-2023-discomat}. Figure~\ref{fig:performance_matsie}c shows that \llamattwo{-Chat} models exhibit the superior performance with results slightly better than the commercial LLMs. Interestingly, in the case of tables, we observe that the performance of commercial LLMs and \llamat{} models are comparable, a distinct feature from previous datasets. The results also confirm a recurring pattern: \llamat{-2} and \llama{-2} models consistently outperform their third-generation counterparts across all evaluation metrics, particularly in chemical label identification and composition extraction tasks. This observation aligns with our previous findings regarding the enhanced domain adaptability of second-generation architectures, suggesting that this advantage extends to structured data interpretation tasks. Detailed performance metrics and task-specific analyses are provided in App.~\ref{appendix:discomat}.

To further analyze the performance, we plot the performance of the best-in-class models for each of the tasks in the dataset in Figure~\ref{fig:performance_matsie}d. In contrast to other datasets, we observe that the superior performance of \llamat{} models are primarily due to regex tables. In other tasks, commercial LLMs exhibit comparable performance to \llamat{} models, sometimes even outperforming them. This suggests that the tabular structure requires special attention. The commercial models, potentially due to the large number of tables potentially present in their pretraining dataset, exhibit superior performance than \llamat{}. However, \llamat{} still outperforms commercial LLMs in regex tables, where complex domain specific notations are used to represent materials, albeit in a tabular fashion.

\subsection{Crystal Generation}
Crystal generation represents a fundamental challenge in materials discovery, traditionally addressed through computationally intensive methods such as density functional theory (DFT) calculations. More recently, generative models~\cite{xie2022crystal, jiao2024crystal, levy2024symmcd, miller2024flowmm}, and Graph Neural Networks ~\cite{lee2023matsciml, duval2023hitchhiker, miret2023the, bihani2024egraffbench} have also been used for generating novel crystal structures. Language models offer an alternative paradigm despite lacking explicit crystallographic optimization. Recent works~\cite{gruver2024fine,antunes_crystal_2024,ding2024matexpert} demonstrate the potential of LLMs toward crystal generation.

We systematically evaluated \llamat{}'s crystal generation capabilities through a comprehensive three-phase optimization strategy for \llamat{-CIF}: CIF pretraining with natural language descriptions, crystallographic instruction finetuning, and PEFT for structure generation. The instruction finetuning phase employed a dual-framework approach comprising syntactic tasks focused on CIF file structure interpretation (e.g., atomic frequency quantification, spatial coordinate analysis, and crystallographic parameter determination) and semantic tasks targeting crystal stability principles (e.g., elemental co-occurrence patterns, atomic spatial distributions, and stability-determining properties). This methodology generated approximately 7 million instruction-output pairs (6,941,865 training instances and 27,183 validation instances), enabling \llamat{} to develop robust comprehension of both CIF file architecture and fundamental materials science principles. The syntactic tasks emphasized structural parameter extraction and validation, while semantic tasks focused on property-conditioned crystal generation and structural prediction, collectively enhancing the model's ability to generate physically meaningful crystal structures.

Quantitative assessment reveals \llamat{-2-CIF}'s exceptional performance across multiple metrics (Tab.\ref{tab:performance_metrics}), achieving near-perfect composition validity (0.995) and coverage (0.986 recall, 0.996 precision), with 49.49\% of generated structures exhibiting thermodynamic stability. The property distribution analysis, quantified through Wasserstein distance measures, demonstrates significant divergence from the training dataset ($\rho$ = 0.623, $N_{el}$ = 0.023), indicating the model's capacity to generate novel structures while maintaining realsitic physical and chemical features. This performance substantially surpasses that of comparably-sized PEFT-finetuned \llama{-2} models, validating our comprehensive domain adaptation strategy. The improvement is particularly evident in structural validity metrics, though model performance exhibits sensitivity to hyperparameter selection\cite{peft_sengupta2024robust} (App.\ref{app:finetune_los}).

Reinforcing our earlier observations of adaptation rigidity across MatNLP and structured information extraction tasks, \llamat{-3}'s crystal generation capabilities follow a similar pattern. While it generates more complex structures (peaks around 24-32 elements versus 6-12 for \llamat{-2-CIF}), it exhibits significantly lower structural validity (0.674) and reduced generation efficiency, requiring approximately 2.5 times more attempts (33,000 versus 13,000) to produce 10,000 evaluation-ready structures. This consistent manifestation of adaptation rigidity across diverse tasks---from natural language processing to crystal structure generation---suggests a fundamental limitation in the domain adaptability of LLMs.

\begin{figure}
    \centering
    \includegraphics[width=0.8\linewidth]{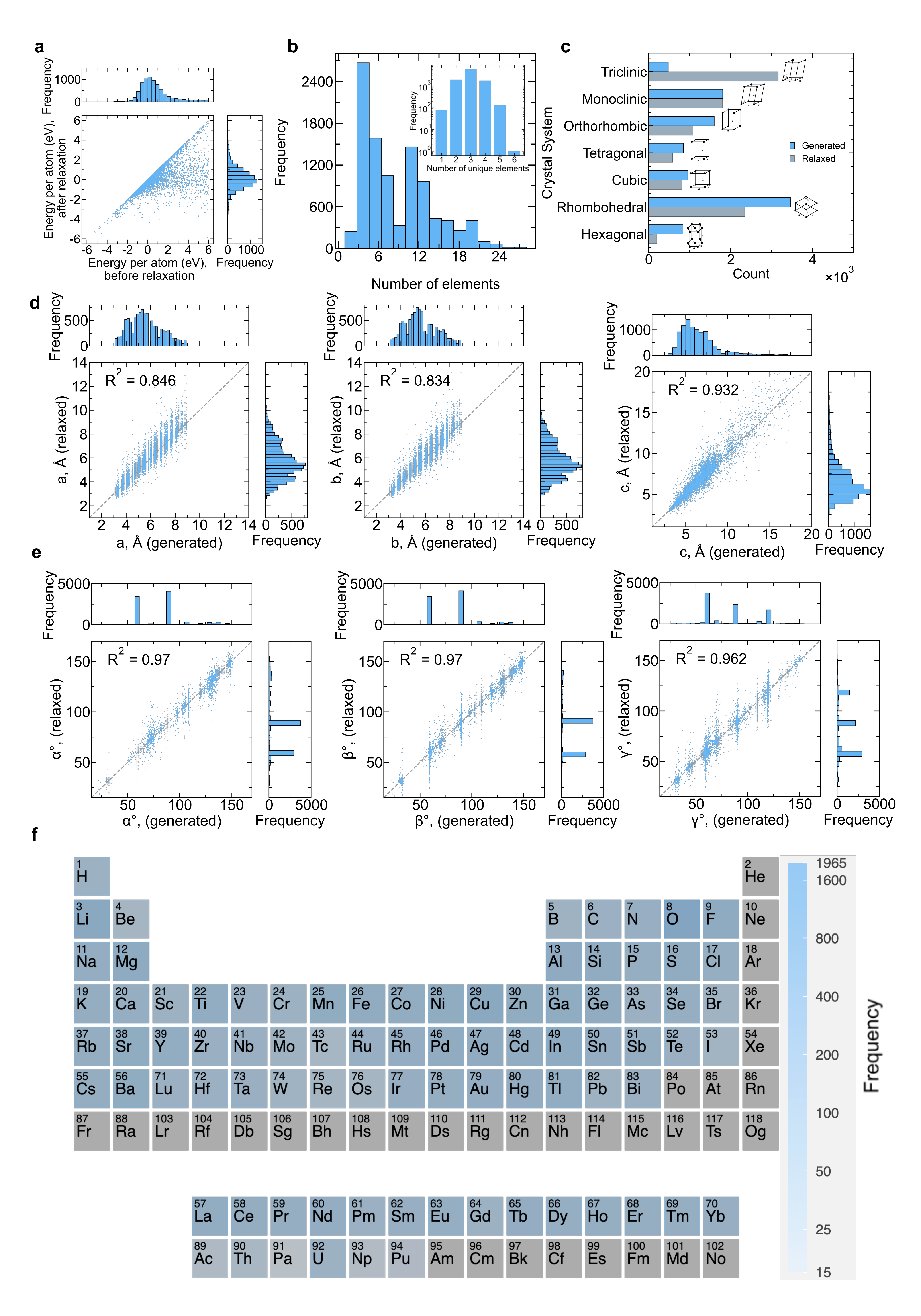}
    \caption{\textbf{Comparative compositional and structural analysis of 10,000 crystal structures generated by \llamat-2-CIF model and their relaxed counterparts.} \textbf{a}, Energy per atom (eV/atom); \textbf{b}, Number of elements in each crystal structure. The inset shows the number of crystals with the unique number of elements; \textbf{c}, The distribution of Bravais lattice systems; \textbf{d}, Lattice parameters (unit cell lengths \textit{a}, \textit{b}, and \textit{c} along x, y, and z-axes; \textbf{e},  Lattice parameters ($\alpha$, $\beta$, and $\gamma$, i.e., the angles between \textit{b} and \textit{c}, \textit{a} and \textit{c}, and \textit{a} and \textit{b}; \textbf{f}, Periodic table heat map visualizing elemental frequency, where color intensity represents generation frequency. Grey cells indicate elements absent in generated structures.}
    \label{fig:cif_performance}
\end{figure}

Detailed analysis of the generated structures reveals distinct characteristics (Fig. \ref{fig:cif_performance}). Note that the results of only \llamattwo{-CIF} is shown in the figure and the corresponding results of \llamatthree{-CIF} is in the Appendix. Energy calculations performed using M3GNet\cite{m3rgmet_chen2022universal}, consistent with previous studies\cite{gruver2024fine}, demonstrate that \llamat{-2-CIF} generates structures predominantly near their ground state energies. This is evidenced by symmetrical, zero-centered energy distributions in both initial and relaxed states (Fig. \ref{fig:cif_performance}\textbf{a}), indicating inherent thermodynamic stability. The compositional landscape exhibits systematic trends, with \llamat{-2-CIF} favoring structures containing 2-4 unique elements and showing exponentially decreasing frequency for higher component counts (Fig. \ref{fig:cif_performance}\textbf{b}). In contrast, \llamat{-3-CIF} generates structures with higher elemental complexity (24-32 elements) compared to \llamat{-2-CIF}'s simpler compositions (6-12 elements), though both models maintain thermodynamic reasonability.

Interestingly, upon relaxation the generated structures exhibited changes in the crystal lattice system (Fig. \ref{fig:cif_performance}\textbf{c}). Initial structures demonstrate a strong preference for rhombohedral symmetry (\textasciitilde4,000 instances), characterized by equivalent lattice parameters. However, relaxation induces a dramatic 6.65-fold increase in triclinic structures, suggesting an inherent tendency toward lower symmetry states. Notably, monoclinic systems exhibit exceptional structural stability post-relaxation. Similarly, lattice parameter analysis (Fig. \ref{fig:cif_performance}\textbf{d,e}) uncovers differential responses to relaxation: unit cell dimensions (\textit{a}, \textit{b}, \textit{c}) show moderate correlations (R$^2$ = 0.83-0.93) between initial and relaxed states, while angular parameters ($\alpha$, $\beta$, $\gamma$) maintain remarkably high correlations (R$^2$ > 0.97). The preservation of characteristic angles (60\textdegree, 90\textdegree, 120\textdegree) indicates retention of fundamental crystallographic motifs despite dimensional adjustments. The elemental composition distribution (Fig. \ref{fig:cif_performance}\textbf{f}) reveals chemical biases aligned with synthetic accessibility: minimal actinide incorporation (\textless50 instances), uniform distribution across transition metals (200-400 instances), and predominant oxygen presence (\textgreater1,600 instances). These patterns reflect both natural abundance and practical synthetic constraints. Note that the \llamat{-CIF} framework demonstrates versatility beyond structure generation, extending to various CIF-related tasks including interpreting the CIF files and corresponding crystal structures (App.~\ref{appendix:cif_prompts}). Altogehter


\begin{table}[ht]
    \centering
    \caption{\textbf{Comparison of crystal structure generation capabilities across different model architectures.} Performance evaluation using multiple metrics: validity (structural integrity and composition correctness), coverage (recall and precision of generated structures), property distribution (Wasserstein distance for density ($\rho$) and number of elements ($N_\text{el}$)), and thermodynamic stability (percentage of structures predicted stable by M3GNet). Arrows indicate metrics' desired direction ($\uparrow$: higher is better, $\downarrow$: lower is better). The top section shows baseline results from state-of-the-art methods~\cite{gruver2024fine}. \llamat{}-2-CIF demonstrates superior performance across most metrics, particularly in composition validity (0.995) and stability prediction (49.49\%), while maintaining high coverage (0.986 recall, 0.996 precision). Bold values indicate the best performance for each metric.}
    \small
    \setlength{\tabcolsep}{4pt}
    \begin{tabular}{@{}lccccccc@{}}
    \toprule
    & \multicolumn{2}{c}{\textbf{Validity}} & \multicolumn{2}{c}{\textbf{Coverage}} & \multicolumn{2}{c}{\textbf{Property Dist.}} & \textbf{Stability} \\
    \cmidrule(lr){2-3} \cmidrule(lr){4-5} \cmidrule(lr){6-7} \cmidrule(lr){8-8}
    \textbf{Method} & Struct.$\uparrow$ & Comp.$\uparrow$ & Recall$\uparrow$ & Prec.$\uparrow$ & $\rho$$\downarrow$ & $N_\text{el}$$\downarrow$ & M3GNet$\uparrow$ \\
    \midrule
    CDVAE~\cite{gruver2024fine}         & \textbf{1.000} & 0.867 & \textbf{0.991} & 0.995 & 0.688 & 1.43 & 28.8\% \\
    \midrule
    \textsc{LLaMA}-2~\cite{gruver2024fine} & & & & & & & \\
    7B ($\tau = 1.0$) & 0.918 & 0.879 & 0.969 & 0.960 & 3.850 & 0.96 & 35.1\% \\
    7B ($\tau = 0.7$) & 0.964 & 0.933 & 0.911 & 0.949 & 3.610 & 1.06 & 35.0\% \\
    13B ($\tau = 1.0$) & 0.933 & 0.900 & 0.946 & 0.988 & 2.200 & 0.05 & 33.4\% \\
    13B ($\tau = 0.7$) & 0.955 & 0.924 & 0.889 & 0.979 & 2.130 & 0.10 & 38.0\% \\
    \midrule
    \textbf{Present work} & & & & & & & \\
    \llamat{-}2-CIF & 0.878 & \textbf{0.995} & 0.986 & \textbf{0.996} & \textbf{0.623} & \textbf{0.023} & \textbf{49.49\%} \\
    \llamat{-}3-CIF & 0.674 & 0.693 & 0.925 & 0.994 & 12.355 & 0.261 & 42.95\% \\
    \bottomrule
    \end{tabular}
    \label{tab:performance_metrics}
\end{table}

\newpage
\section{Discussion}
\vspace{-0.1in}
LLMs have revolutionized several fields including materials. However, applications of LLMs to scientific domains require their adaptation ensuring reliability, superior performance, and possibility for large-scale deployment without excessive computational and economic overhead. Through \llamat{}, we demonstrate that domain-adapted foundational language models for materials can outperform significantly larger open-source LLMs. A comprehensive evaluation of \llamat{} on several tasks, including entity recognition, entity extraction, and information extraction from text and tables, demonstrates that \llamat{} models significantly outperform general purpose LLMs such as GPT, Claude, and Gemini. Thus, strategic domain adaptation through CPT and targeted IFT can transform LLMs into specialized scientific tools without compromising their foundational capabilities. The fact that the present work relies on the smaller models of \llama{} family suggests that adapting smaller models toward a specific domain might be a more economical and practical solution than relying on general-purpose LLMs.


The \llamat{-CIF} models represent a particularly significant advance in materials structure prediction. Generative modeling of crystals is a challenging task widely explored using several methods. Our results demonstrate that domain-adapted LLMs can be powerful tool for generative modeling of crystals. The models' demonstrated ability to implicitly learn realistic chemical constraints—evidenced by systematic trends in elemental compositions and crystal system preferences—suggests potential for accelerating materials discovery while maintaining physical and chemical validity. Further, the textual nature of LLMs could potentially be exploited further to explore the synthesis pathways to realize the generated crystals. 

A significant finding emerges in the differential performance between model generations. Despite \llamatthree{}'s superior baseline capabilities, \llamattwo{} variants demonstrate enhanced adaptability across multiple tasks, particularly in tabular information extraction and crystal structure generation. This raises an interesting question about the ability of highly over-trained models, such as \llamathree{} to adapt to a new domain through CPT~\cite{rigid_1,rigid_2}. This observation, referred to as ``adaptation rigidity'', reported for the first time to the best of the authors' knowledge, challenges the conventional scaling assumptions in LLMs. We hypothesize that the loss landscape\cite{loss} in the local vicinity of the minima in over-trained ~\llamathree{} models may have a notably different character in comparison to those of \llamatwo{}. This observation suggests the need to reevaluate scaling strategies in domain-specific AI applications, potentially influencing the development trajectory of specialized language models across scientific domains. 


Looking ahead, this work establishes a foundation for integrating AI systems into materials research workflows. The demonstrated capabilities in automated literature analysis, extraction, and crystal structure prediction suggest the potential for accelerating materials discovery pipelines~\cite{miret2024llms}. Future development should focus on enhancing model robustness, expanding capabilities to broader MatSci applications, and developing theoretical frameworks for understanding domain adaptation in LLMs. The insights gained from this study—toward developing foundational LLMs for materials—may inform fundamental principles for developing specialized AI systems across scientific domains, potentially transforming how we approach the use of LLMs for scientific applications.

\newpage
\section{Methods}
\label{sec:Methodology}
\vspace{-2mm}
\subsection{Dataset Preparation}
\vspace{-2mm}
\subsubsection{Pretraining Dataset: R2CID}
The performance of foundation models is fundamentally determined by their pretraining dataset composition, necessitating meticulous curation of the constituent data sources. Our pretraining dataset, designated R2CID, integrates three distinct components: scientific literature from materials research publications, a curated subset of RedPajama (the original pretraining corpus for \llama{} models), and crystallographic information files (CIF). The scientific literature provides comprehensive materials characterization and synthesis protocols, while the RedPajama subset help prevent the catastrophic forgetting of the English language processing capabilities. The CIF datasets provide information on crystal structures, including atomic positions, lattice parameters, and symmetry operations. This tripartite combination enabled continued pretraining to generate the \llamat{} models. The specific composition and characteristics of each dataset component are detailed below.

\label{headings} 
\textbf{a. Research Papers.}
Our corpus comprises over 4 million peer-reviewed articles sourced from approximately 500 Elsevier~\cite{elsevier} and 300 Springer~\cite{springer} journals. Selection criteria included full-text accessibility in XML format for Elsevier publications and HTML format for Springer publications. Journal selection was made manually based on the relevance to the materials domain. Article acquisition utilized the CrossRef API~\cite{crossref} to extract Digital Object Identifiers (DOIs), facilitating subsequent retrieval of full-text content in publisher-specific formats.

\textbf{b. RedPajama.}
The RedPajama dataset~\cite{redpajama}, which served as the primary training corpus for the \llamattwo{}~\cite{llama2}, encompasses diverse textual sources, including arXiv preprints, GitHub repositories, StackExchange discussions, Wikipedia articles, and sanitized Common Crawl data. To preserve the model's foundational linguistic capabilities while preventing catastrophic forgetting, we extracted a representative subset of approximately 700 million tokens. This strategic sampling maintains the model's general-purpose functionality while facilitating domain-specific knowledge acquisition.

\textbf{c. Crystallographic Information Files.}
Despite the existence of multiple text-based crystal representations~\cite{alampara2024mattext}, crystallographic information files (CIF) remain the definitive standard for structural data derived from diffraction studies. These standardized files encode essential parameters, including unit cell dimensions, interaxial angles, space group symmetry operations, and atomic position coordinates. Our dataset incorporates 470,000 CIF files, augmented with natural language descriptions generated via RoboCrystallographer~\cite{Ganose_Jain_2019}. These files were aggregated from three major sources: the Materials Project~\cite{materials_project}, GNoME-based ab-initio configurations~\cite{gnome}, and the American mineralogist crystal structure database (AMCSD)~\cite{amcsd}.

\textbf{d. R2CID Dataset Integration.}
The integration protocol implemented a structured mixing strategy to optimize training efficiency and maintain model robustness. Research paper content was systematically interspersed with RedPajama text, maintaining a ratio of 2.4 million RedPajama tokens per 100 million research paper tokens. Crystallographic data integration occurred within the terminal 10\% of the dataset, where CIF files and their descriptions were interleaved with research paper content.
\subsubsection{Instruction Finetuning}  \label{appendix:ift}
The IFT protocol incorporated multiple specialized datasets encompassing materials science and general question-answering tasks. We developed two novel domain-specific datasets: MatBookQA, consisting of materials science questions and answers generated via GPT4 using contextual prompting, and a comprehensive question bank derived from the Graduate Aptitude Test in Engineering (GATE). GATE is a standardized examination for postgraduate admissions at premier Indian institutions and select international universities. The constituent datasets are detailed below.

\textbf{a. OpenOrca.}  \label{appendix:orca}
The OpenOrca corpus encompasses 800,000 high-fidelity instruction-response pairs spanning diverse technical domains. Previous investigations~\cite{mukherjee2023orca} have demonstrated that models finetuned on this dataset exhibit superior performance across multiple evaluation frameworks, including Big-Bench Hard and AGIEval. This enhanced performance manifests in improved technical comprehension, complex query resolution, and domain-appropriate response generation. Dataset optimization procedures were implemented to determine the optimal training sample size for our specific application (see App.~\ref{appendix:section:dataset_optimization}).

\textbf{b. Mathematics Corpus (MathQA).} \label{appendix:math}
To enhance the model's quantitative reasoning capabilities, we incorporated 7,500 selected problems from the Math dataset~\cite{math}. This curated subset consists of advanced competition-level mathematical problems chosen to develop robust problem-solving abilities across various mathematical domains.

\textbf{c. Materials Science Instruction Sets (MatSciInstruct).} \label{appendix:matsciinstruct}
The materials science instruction corpus integrates multiple specialized datasets, including a novel collection generated through GPT-4 (gpt-4-0613) using open-source materials science textbooks as source material. This approach generated contextually rich questions spanning diverse materials science subdomains. The corpus incorporates MatSciInstruct\cite{song-etal-2023-honeybee}, which employs a two-phase development framework: an initial Generation phase utilizing an instructor model to create domain-specific instruction data, followed by a Verification phase wherein a distinct verifier model assesses instruction quality across multiple dimensions including accuracy, relevance, completeness, and logical consistency. The instruction set is further augmented with the MatSciNLP training corpus and our custom-developed MatBookQA dataset.

\textbf{d. MatBookQA.} \label{appendix:matbookqa}
The MatBookQA dataset was systematically developed using a comprehensive materials science textbook\cite{matbook_Hutagalung_2012}. The development protocol employed chapter-wise GPT-4 prompting using twenty distinct prompt templates (detailed in Appendix~\ref{appendix:prompts_matbookqa}), equally divided between generating short and extended responses. This methodology yielded 2,069 question-answer pairs, comprising 1,887 concise responses and 182 comprehensive explanations.

\textbf{e. Materials Science Question Answering (MaScQA).} \label{appendix:mascqa}
The MaScQA dataset encompasses 1,585 questions from Indian undergraduate engineering examinations, specifically 1,036 from civil engineering and 549 from chemical engineering curricula. Answer validation was performed using the GPT-4o model (2024-02-01), with only verified correct responses retained in the final dataset. As detailed in Zaki et al.\cite{zaki_mascqa}, the question taxonomy includes four distinct categories: traditional multiple-choice, correlation-based matching, numerical multiple-choice, and open-ended numerical problems.

\textbf{f. Crystallographic Information File (CIF) Dataset.} \label{ift:cif}
To train the language models to generate crystals, we created a new set of tasks that enable the language models to train on various aspects of CIF. Specifically, we developed instruction-output pairs from CIF files sourced from AMCSD, Google GNoME, and the Materials Project to enhance \llamat's crystallographic comprehension and natural language query resolution capabilities. To this extent, we developed an instruction set implementing a dual-task framework comprising syntactic and semantic components. Syntactic tasks address the structural interpretation of CIF files. In contrast, semantic tasks, inspired by Gruver et al. (2024)\cite{gruver2024fine}, focus on crystal stability principles, including elemental co-occurrence patterns, atomic spatial distributions, and stability-determining properties. This methodology generated approximately 7 million instruction-output pairs (6,941,865 training instances and 27,183 validation instances). The complete task framework, with corresponding system prompts detailed in Appendix~\ref{appendix:cif_prompts}, encompasses:

\textbf{Syntactic Analysis Tasks:}
\begin{itemize}
\item Atomic frequency quantification within crystal structures.
\item Spatial coordinate-based atomic identification.
\item Crystal parameter determination: dimensional analysis, volumetric calculation, and space group classification.
\item Site occupancy equivalence evaluation.
\item Structure-based chemical formula derivation.
\end{itemize}
\textbf{Semantic Analysis Tasks:}
\begin{itemize}
\item Property-conditioned crystal structure generation.
\item Positional atomic prediction using \texttt{MASK} token methodology.
\item Structural dimension prediction for stability optimization.
\item Element-constrained crystal structure synthesis.
\end{itemize}
\subsubsection{Materials Natural Language Processing (MatNLP)}
The model evaluation employed a comprehensive dual-stage assessment protocol encompassing both materials science and general language capabilities. The primary evaluation phase compared multiple model iterations to optimize architectural decisions, while the secondary phase benchmarked performance against contemporary state-of-the-art materials science models. The primary evaluation corpus comprised 14 specialized materials science tasks, supplemented with four general-purpose reasoning and comprehension assessments to preserve broad linguistic capabilities.

Table~\ref{app:downstream_data} and App.~\ref{appendix:task_category_description} delineate the task taxonomy, dataset specifications, and sample distribution across training and validation sets. The evaluation framework encompasses multiple task categories, namely, sentence classification (SC), relation extraction (RE), named entity extraction (NER), synthesis action retrieval (SAR), paragraph classification (PC), entity extraction (EE), slot filling (SF), question answering (Q\&A), and multiple choice question answering (MCQ). Detailed task specifications are documented in App.~\ref{appendix:task_category_description} and Ref.~\cite{song-etal-2023-matsci}. Model evaluation incorporated single-epoch fine-tuning on the training corpus prior to validation assessment to ensure instruction comprehension.

The secondary evaluation phase utilized the MatSciNLP dataset~\cite{song2023matsci}, which reformulates these tasks as multi-class classification problems. This meta-dataset enables direct performance comparison with existing materials science language models. To maintain evaluation integrity, distinct model instances were trained for each evaluation phase due to potential dataset overlap. Performance assessment followed the methodology established in Ref.~\cite{song-etal-2023-honeybee}, implementing single-epoch training on a condensed training set followed by evaluation on a comprehensive 170,000 sample validation corpus. Task-specific examples are provided in Appendix~\ref{appendix:examples}.

\subsubsection{Structured Information Extraction Dataset (MatSIE)}

The extraction of structured information facilitates automated data processing and machine-readable format conversion. Given the domain expertise and structured data comprehension acquired through instruction fine-tuning, \llamat{} models were hypothesized to demonstrate robust performance in structured extraction tasks. To further analyze this capability, we performed the evaluation of the models using instruction-output pairs derived from four specialized datasets: (i) Doping, (ii) General materials, (iii) metal-organic frameworks (MOF)~\cite{nerre}, and (iv) \textsc{{DiSCoMaT}}~\cite{gupta-etal-2023-discomat}. 

The initial three datasets focus on entity recognition and relationship extraction within materials science texts. The \textsc{DiSCoMaT} dataset provides annotated tables extracted from materials science publications. For the entity-relationship datasets, we developed six system prompts serving as prefixes to query-response pairs, where responses conform to standardized JSON schemas as established in Ref.~\cite{nerre} (see App.~\ref{appendix:nerre}). The \textsc{DiSCoMaT} dataset, originally developed for alternative applications, was transformed to generate JSON-structured annotations suitable for the language models (format specifications in App.~\ref{appendix:discomat}).
\subsubsection{Evaluation}
Evaluation for Doping, MOFs, and General materials tasks were done in the same way as mentioned in \cite{nerre}, but the evaluations were only performed on the outputs that could be parsed as json. No manual human evaluation was done. Also note that the tasks in these datasets had less support compared to the MatNLP tasks, as can be seen in \ref{appendix:task_descriptions}.
Evaluations on \textsc{DiSCoMaT} dataset was also evaluated similarly, with only outputs parsed as json format were considered. We calculated accuracy for each task separately, and the exact match criteria counts how many outputs matched exactly with the gold answers. 
\subsection{Model Development Methodology}
\subsubsection{Continued Pretraining}
The pretraining corpus underwent hierarchical prioritization based on materials science relevance (P1 > P2 > P3). This corpus integrated materials science community discourse data and incorporated RedPajama subset to mitigate catastrophic forgetting, supplemented with 470,000 crystallographic information files for structural comprehension. The integration methodology employed a dual-phase mixing strategy:
\begin{itemize}
    \item Primary phase: 90\% of P1 content integrated with P2 and P3 datasets through stochastic shuffling.
    \item Secondary phase: Remaining 10\% of P1 content combined with the CIF dataset through stochastic shuffling.
\end{itemize}

The resultant dataset underwent final integration with RedPajama using a token-ratio methodology: approximately 0.15M RedPajama tokens per 5M materials science tokens. The details of the pretraining dataset, along with the number of tokens, are mentioned in \ref{tab:cpt_data}.

\subsubsection{Instruction Finetuning}
\label{method:ift}
The \llamat-Chat models, initialized with corresponding \llamat\ model weights, underwent tri-phase instruction finetuning:
\begin{itemize}
\item \textbf{Phase I:} Single-epoch finetuning on \hyperref[appendix:orca]{OpenOrca} dataset to establish general instruction-following capabilities
\item \textbf{Phase II:} Three-epoch finetuning on \hyperref[appendix:math]{mathematical questions}, optimizing quantitative reasoning capabilities. The limited dataset size enabled the observation of continuous validation loss reduction.
\item \textbf{Phase III:} Single-epoch finetuning on an integrated corpus comprising \hyperref[appendix:matsciinstruct]{MatSciInstruct, MatSciNLP}, \hyperref[appendix:matbookqa]{MatBookQA}, and \hyperref[appendix:mascqa]{MaScQA}, focusing on materials science-specific instruction comprehension
\end{itemize}

Implementation utilized the Megatron-LLM framework with learning rate initialization at $2 \times 10^{-6}$, scaling to $2 \times 10^{-5}$ over initial 10\% iterations, followed by cosine decay. This protocol was replicated for \llamattwo and \llamatthree chat model development.

\subsubsection{Task Finetuning}
\textbf{a. \llamat{-Chat}.} The final development phase incorporated combined training on the training set of MatNLP and MatSIE datasets. This phase employed a $10^{-5}$ learning rate with cosine decay over two epochs. The intention of this stage was to familiarize the \llamat{-Chat} models with a wide range of tasks relevant to materials research, including scientific natural language processing, structured information extraction, and tabular information extraction. All the training data from these datasets were mixed to form a single task dataset on which the \llamat{-Chat} models were finetuned.

\textbf{b. \llamat{-CIF}.} Crystal structure generation capabilities were implemented through parameter-efficient finetuning~\cite{gruver2024fine}. Optimal \llamat\ checkpoints underwent instruction finetuning using the dataset detailed in Section~\ref{appendix:ift}, with model selection based on minimal validation loss (Fig.~\ref{fig:cif_loss}). Comprehensive finetuning specifications and hardware configurations are documented in Sections~\ref{app:finetune_los} and ~\ref{appendix:hardware}.

\subsection{Baselines}
In order to compare the performance of \llamat{} with existing general-purpose models, we considered \llama{}, Gemini-1.5 Flash-8B, and Claude-3 Haiku. Note that these models were chosen as they were the closest comparable ones in the respective families with \llamat{} models in terms of the number of parameters. To assess the effect of finetuning, \llama{} models were evaluated both with and without finetuning (FT). \\

\subsection{Evaluation Metrics}
\textbf{a. Loss function.} The loss function used to train the models for CPT, IFT, and task finetuning is the cross-entropy loss. 

\textbf{b. MatNLP and MatSIE.} To evaluate the performance of models on the downstream tasks in MatNLP and MatSIE, precision, recall, and F1 scores are used with the annotated data as the ground truth.

\textbf{c. Crystal generation.} 
To evaluate the performance of LLMs for crystal generation, we rely on the following metrics.
\begin{enumerate}
    \item Validity check: Structural validity and compositional validity are calculated as described in \cite{xie2022crystal}. The former indicates that the distance between the centres of two atoms is greater than the sum of their atomic radii. The compositional validity is obtained using SMACT\cite{validity-2_davies2019smact}, which identifies if the given material is charge neutral based on all possible charge combinations.
    \item Coverage: We use two coverage metrics, COV-R (recall) and COV-P (precision), described in \cite{xie2022crystal} to measure the similarity between ensembles of generated materials and ground truth materials in the test set. COV-R Measures the percentage of ground truth materials being correctly predicted, and COV-P measures the percentage of predicted materials having high quality as described in \cite{xie2022crystal}
    \item Property statistics: We compute the Wasserstein distance between the property distributions of the generated materials and the test materials. We use density (in g/ cm$^{3}$) and number of unique elements ( \#elem) as the properties.
    \item Stability Check: We used M3GNet (\cite{chen2022universal}) to approximate force, energy, and stress in crystal unit cells. We use the predicted energy of the final structure as our stability metric since those having low predicted absolute energy ( < 0.1 eV/atom $\hat{E}_{hull}$) are likely to be stable. While other potentials could be used, we relied on M3GNet to ensure direct comparison with the baselines.
\end{enumerate}

\section{Code availability}
Codes used in this work are shared in the \href{https://github.com/M3RG-IITD/llamat}{\llamat{}} GitHub repository: \href{https://github.com/M3RG-IITD/llamat}{https://github.com/M3RG-IITD/llamat}.

\section*{Acknowledgments}
N. M. A. K. acknowledges the funding support received from BRNS YSRA (53/20/01/2021-BRNS), ISRO RESPOND as part of the STC at IIT Delhi, Google Research Scholar Award, Intel Labs, and Alexander von Humboldt Foundation. M. acknowledges grants by Google, IBM, Microsoft, Wipro, and a Jai Gupta Chair Fellowship. M. Z. acknowledges the funding received from the PMRF award by the Ministry of Education, Government of India. The authors thank Microsoft Accelerate Foundation Models Research (AFMR) for access to OpenAI models. The authors thank the High-Performance Computing (HPC) facility at IIT Delhi for computational and storage resources. This work was partially supported by the Edinburgh International Data Facility (EIDF) and the Data-Driven Innovation Programme at the University of Edinburgh. The EIDF provided access to Cerebras CS2 clusters for training the language models.
\newpage
\bibliographystyle{unsrt}  
\bibliography{references}  

\newpage
\appendix
\section*{Appendices}
\counterwithin{figure}{section}
\counterwithin{table}{section}

\section{Dataset details}
\label{app:dataset}
\subsection{Pretraining and IFT dataset}

Table \ref{appendix:ift_pretrain_datasets} contains details about the datasets we used for pretraining, followed by instruction finetuning to infuse the materials domain knowledge to the model while also giving our model the capability to follow instructions and answer queries through chat.

\begin{table}[H]
\centering
\caption{details about Instruction finetuning and pretraining datasets. for more detailed info, see Sec.~\ref{method:ift}}
\label{appendix:ift_pretrain_datasets}
\renewcommand{\arraystretch}{1.2} 
\scalebox{0.7}{
\begin{tabular}{lllp{10cm}}
\toprule
\textbf{Pretraining Dataset} & \textbf{Token Length} & & \\
\midrule
Elsevier/Springer & 30B & - & Tokens sourced from material science research papers on Elsevier and Springer. \\
RedPajama & 300M & - & A part of the Original Llama-2 corpus. We interleave this at regular intervals in the pretraining corpus: 10M research paper tokens followed by 0.1M RedPajama tokens. \\
\shortstack{Mat Sci Community \\  Discourse} & 30M & - & Tokens sourced from MSCD, which is a forum for questions and answers for material science. \\
\midrule
\textbf{IFT Dataset} & \textbf{Train Size} & \textbf{Val Size} & \textbf{Description} \\
\midrule
OpenOrca & 576,000 & - & The standard instruction finetuning dataset. A subset of the FLAN dataset is augmented with answers from GPT-4. It contains generic instructions following tasks. \\
MathQA & 7500 & 5000 & Contains numerical math questions. We train on this dataset to improve the mathematical ability of our model. \\
\midrule
MatSciInstruct & 52658 & - & A collection of NLP tasks in the material science domain generated using ChatGPT, Claude, and GPT-4\cite{song-etal-2023-honeybee}\\
MatSciNLP & 19942 & 170594 & A collection of NLP tasks in the material science domain\\
MatBookQA & 150 + 1800 & 32 + 87 & Long and short questions and answers generated by GPT-4 on chapters of an open-source material science book. \\
MaScQA $\times$ 4 & 1022 $\times$ 4 & 1022 $\times$ 4 &  comprises 1036 and 549 questions from the civil and chemical engineering undergraduate-level exams in India, respectively. Only the questions answered correctly by GPT4 are taken, making the total count of 1022.\\
\midrule
Crystal finetuning Dataset & 6,941,865 & 27,183 & Semantic and syntactic instruction-output pairs based on CIF files. Details are provided in Appendix \ref{appendix:cif_prompts} \\

\bottomrule
\end{tabular}}
\end{table}

\subsection{Pre-processing and tokenization}
Research papers contain in-text references to tables, figures, and related research papers. Therefore, we adopted the pre-processing methodology introduced by Chen et al. (2023)\cite{meditron, meditron_github_epfmedtrn} and replaced them with the respective captions and the citations provided in the research paper. This additional text is added between specific tokens; for example, [FIG\_REF] and [\textbackslash FIG\_REF] were used for providing captions of figures and tables, and [BIB\_REF] and [\textbackslash BIB\_REF] are used for inserting bibliography, respectively.

The Meditron's GitHub repository supported the training of \llamatwo\ models on NVIDIA GPUS. Hence, we used their codebase to tokenize the data and perform the pretraining and fine-tuning experiments on NVIDIA A100 80GB GPUs.

For continued pretraining of \llamathree\ , the dataset was tokenized using the resources provided in the \href{https://github.com/Cerebras/modelzoo/tree/main/src/cerebras/modelzoo/data_preparation/data_preprocessing}{CEREBRAS MODELZOO GitHub repository} and the example of the same is provided \href{https://training-docs.cerebras.ai/getting-started/pre-train-your-first-model}{here}.

Since the support for tokenizer to train \llamathree\ is not provided in \cite{meditron_github_epfmedtrn}, we implemented the \href{https://github.com/meta-llama/llama3/blob/main/llama/tokenizer.py}{TikToken tokenizer} in the Meditron codebase for training the \llamatthree{} chat model and fine-tuning on downstream tasks. Note that \href{https://github.com/meta-llama/llama/blob/main/llama/tokenizer.py}{sentencepiece} and \href{https://github.com/meta-llama/llama3/blob/main/llama/tokenizer.py}{tiktoken} are used in \llamatwo\ and \llamathree\ as tokenizers respectively.

While preprocessing the text for the chat model, each data instance has text corresponding to three roles: system, question, and answer. To inform the model about the same, we use the following template for both the models: \texttt{f"\textless|im\_start|>\{\textbackslash role\}\textbackslash n\{\textbackslash text\}\textless|im\_end|>\textbackslash n"}. This template requires adding two new tokens, i.e., \texttt{\textless|im\_start|>} and \texttt{\textless|im\_end|>} .

\newpage
\subsection{MatNLP, MatSIE, and Crystal generation datasets}
\label{app:downstream_data}
Table \ref{appendix:task_descriptions} contains details about the individual datasets and tasks used for training and evaluating the models. Figure ~\ref{fig:cif_loss} shows the distribution of Bravais lattice on the CIF dataset used to train \llamat{}.

\begin{figure}[H]
    \centering
    \includegraphics[width=0.30\linewidth]{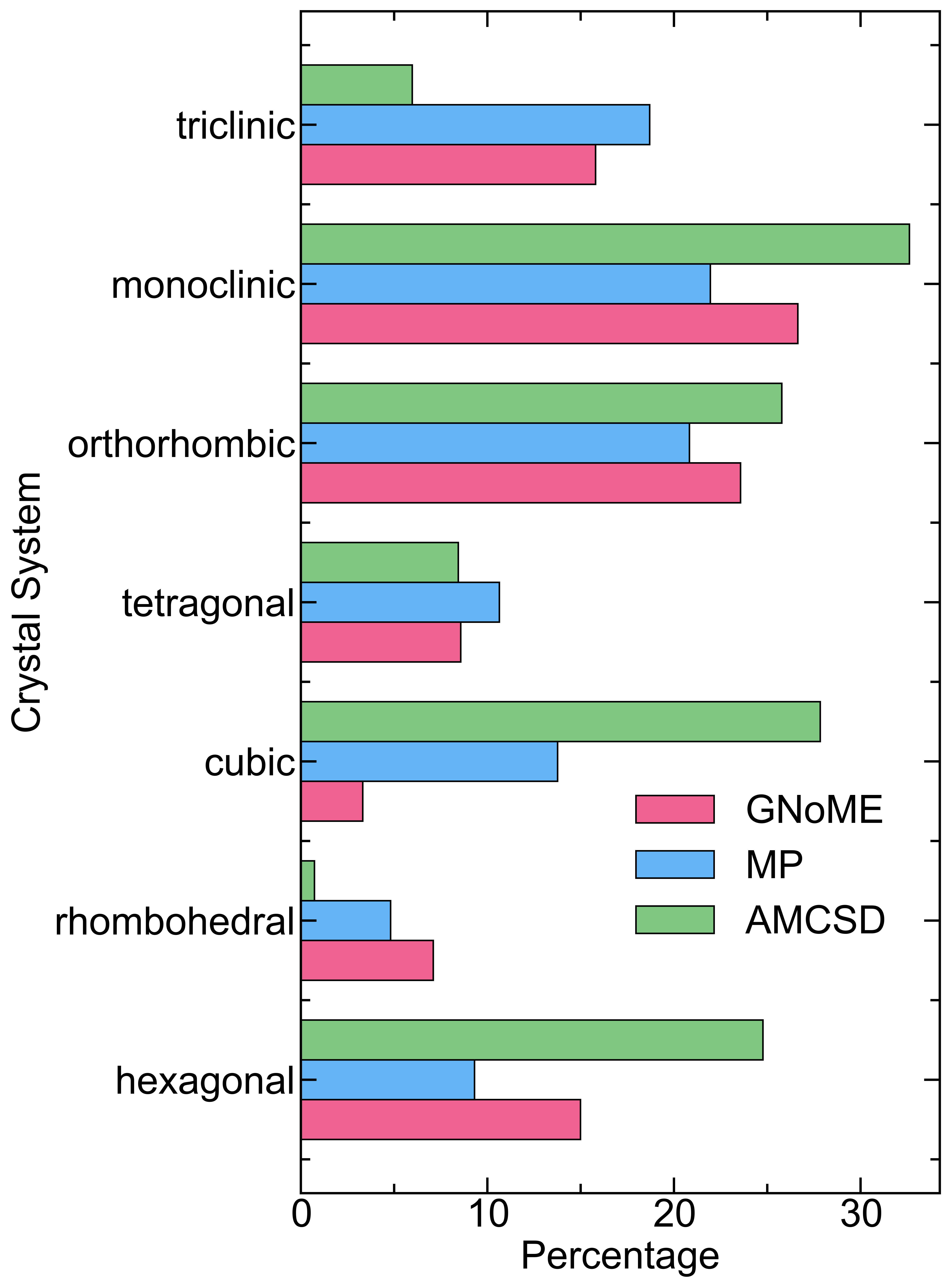}
    \caption{Distribution of the Bravais lattice of CIF training dataset.}
    \label{fig:cif_loss}
\end{figure}


\begin{table}[H]
\centering
\caption{Task Descriptions and evaluation dataset sizes. For a detailed description of each task type, see \ref{appendix:task_category_description}}
\scalebox{0.7}{
\begin{tabular}{llllp{12cm}}
\toprule
\textbf{Task} & \textbf{Dataset} & \textbf{Train Size} & \textbf{Eval Size} & \textbf{Task Description} \\
\midrule
\textbf{MatNLP} & & & & \\
\midrule
\textbf{Entity recognition} & & & & \\
Matscholar & Matscholar & 1062 & 1061 & Named entity recognition tasks over data taken from matscholar. \\
SOFC-1 & sofc-token & 175 & 177 & Named entity recognition over sentences from a corpus with data pertaining to "solid oxide fuel cells" \cite{friedrich2020sofc} \\
SOFC-2 & sofc-token  & 175 & 179 & Identify slot fillers from sentences using a predefined set of semantically meaningful entities. Each sentence describes an experiment frame. \\
SC-CoMIcs-1 & sc-comics & 937 & 936 & Named entity recognition over sentences from a corpus on "superconductivity" \cite{yamaguchi2020sc}. \
\\
\textbf{Classification} & & & & \\
Glass & glass-non-glass & 300 & 299 & Paragraph classification: Determine whether a given paragraph pertains to glass science. This task is adapted from \cite{venugopal2021looking} \\
Synthesis Actions & SAR & 565 & 569 & Classify word tokens into one of eight predefined synthesis action categories. SAR data adapted from \cite{wang2022ulsa}\\
SOFC-3 & sofc-sent & 1893 & 1889 & Sentence classification: Identify sentences that describe relevant experimental facts. The task data is adapted from \cite{friedrich2020sofc} \\
\textbf{Extraction} & & & & \\
SC-CoMIcs-2 & sc-comics & 287 & 288 & Extract event arguments and their roles based on specified event triggers. \\
SC-CoMIcs-3 & sc-comics & 376 & 373 & Predict the most relevant relation type for a given span pair. \\
MatSci & structured-re & 1788 & 1786 & Predict the most relevant relation type for a given span pair. \\
\midrule
\textbf{English} & & & & \\
\midrule
Q\&A & squad & 1042 & 1042 & English questions and answers based on reading comprehension. \\
MCQ & hellaswag & 981 & 980 & English tasks on multiple choice question answering based on common sense. \\
MCQ & boolqa & 500 & 499 & Dataset with naturally occurring yes/no questions. \\
MCQ & story-cloze & 500 & 501 & MCQ for common-sense evaluation for story understanding and generation. Choose the correct ending for a 4-sentence story. \\
\midrule
\textbf{SIE Doping} & & & & \\
\midrule
NER & basemats & 322 & 59 & Entity recognition of the base material used in a sentence referencing the use of doping. \\
NER & dopants & 385 & 66 & Entity recognition of the dopant used in a sentence referencing the use of doping. \\
RE & triplets & 327 & 62 & Relation extraction between base materials and dopants. \\

\midrule
\textbf{SIE General} & & & & \\
\midrule
NER & acronym & 45 & 13 & Entity recognition of the acronym for a material used in the input. \\
NER & applications & 443 & 53 & Entity recognition of the applications for material in the input. \\
NER & name & 216 & 34 & Entity recognition of the name of a material in the input. \\
NER & formula & 417 & 63 & Entity recognition of the formula of a material in the input. \\
NER & structure or phase & 325 & 47 & Entity recognition of the structure or phase of a material in the input. \\
NER & description & 358 & 49 & Entity recognition of the description of a material in the input. \\
RE & formula-name & 103 & 8 & Relation extraction to get which formula corresponds to which material name in the input. \\
RE & formula-structure/phase & 427 & 52 & Relation extraction to get which material formula corresponds to which structure/phase description in the input. \\
RE & formula-application & 811 & 56 & Relation extraction to get which material formula in the input corresponds to which applications. \\
RE & formula-description & 399 & 41 & Relation extraction to get which material formula in the input corresponds to which description. \\

\midrule
\textbf{SIE MOFs} & & & & \\
\midrule
NER & name of MOF & 511 & 65 & Entity recognition of the name for a MOF material in the input. \\
NER & MOF formula & 100 & 16 & Entity recognition of a MOF formula for a material in the input. \\
NER & MOF description & 267 & 22 & Entity recognition of description for a MOF material in the input. \\
NER & guest species & 201 & 26 & Entity recognition of guest species for MOF material mentioned in the input. \\
NER & applications & 1024 & 128 & Entity recognition of applications for a MOF material mentioned in the input. \\
RE & name-guest species & 255 & 34 & Relation extraction of name and guest species mentioned in the input. \\
RE & name-application & 1004 & 137 & Relation extraction of name and applications mentioned in the input. \\
RE & name-description & 168 & 16 & Relation extraction of name and description mentioned in the input. \\

\midrule
\textbf{DiSCoMaT} & & & &   \\
\midrule
Table & comptable & | & | & Detect whether the input table has material compositions. \\
Table & regex & | & | & Detect whether compositions are extractable using a regular expression parser. \\
Table & gid & 5146 & 737 & Detect which column/row is a material identifier present in. \\
Table & composition & | & | & Identify all columns/rows containing complete material composition information. \\
Table & chemical & | & | & Identify all columns/rows reporting values of constituent chemicals of the material. \\
\bottomrule
\end{tabular}}
\label{appendix:task_descriptions}
\end{table}

\section{Task category description}
\label{appendix:task_category_description}
\vspace{-0.1in}
\begin{table}[H]
\centering
\caption{Descriptions of NLP tasks in the MatNLP dataset, with task data adapted from various sources~\cite{song-etal-2023-matsci}}
\renewcommand{\arraystretch}{1.3} 
\begin{tabular}{p{4cm}|p{11cm}}
\hline
\textbf{Task Type} & \textbf{Description} \\
\hline
Named Entity Recognition (NER) &  The  NER task requires models to extract summary-level information from materials science text and recognize entities, including materials, descriptors, material properties, and applications, among others. Identify the best entity type label for a given text span, including handling non-entity spans with a “null” label. NER task data in downstream tasks is adapted from \cite{weston2019named, friedrich2020sofc, mysore2019materials, yamaguchi2020sc} \\
\hline
Relation Extraction (RE) & Predict the most relevant relation type for a given span pair (e.g., $s_i$, $s_j$). MatSci-NLP contains relation classification task data adapted from \cite{mysore2019materials, yamaguchi2020sc, mullick2024matscire}. \\
\hline
Event Argument Extraction (EE) & Extract event arguments and their roles based on specified event triggers, accounting for potential multiple events in a given text. MatSci-NLP task data is adapted from \cite{mysore2019materials, yamaguchi2020sc} \\
\hline
Paragraph Classification (PC) & Determine whether a given paragraph pertains to glass science. This task is adapted from \cite{venugopal2021looking}\\
\hline
Synthesis Action Retrieval (SAR) & Classify word tokens into one of eight predefined synthesis action categories. SAR data in MatSci-NLP is adapted from \cite{wang2022ulsa} \\
\hline
Sentence Classification (SC) & Identifying sentences that describe relevant experimental facts. The task data is adapted from \cite{friedrich2020sofc} \\
\hline
Slot Filling (SF) & Extract slot fillers from sentences using a predefined set of semantically meaningful entities. Each sentence describes an experiment frame, and the model predicts slots for that frame. Task data is adapted from \cite{friedrich2020sofc} \\
\hline
\end{tabular}
\label{tab:tasks}
\end{table}

\section{Hyperparameter optimization}
\label{app:hyperparameters}

\subsection{Pretraining}

\begin{figure}
    \centering
    \includegraphics[width=0.5\linewidth]{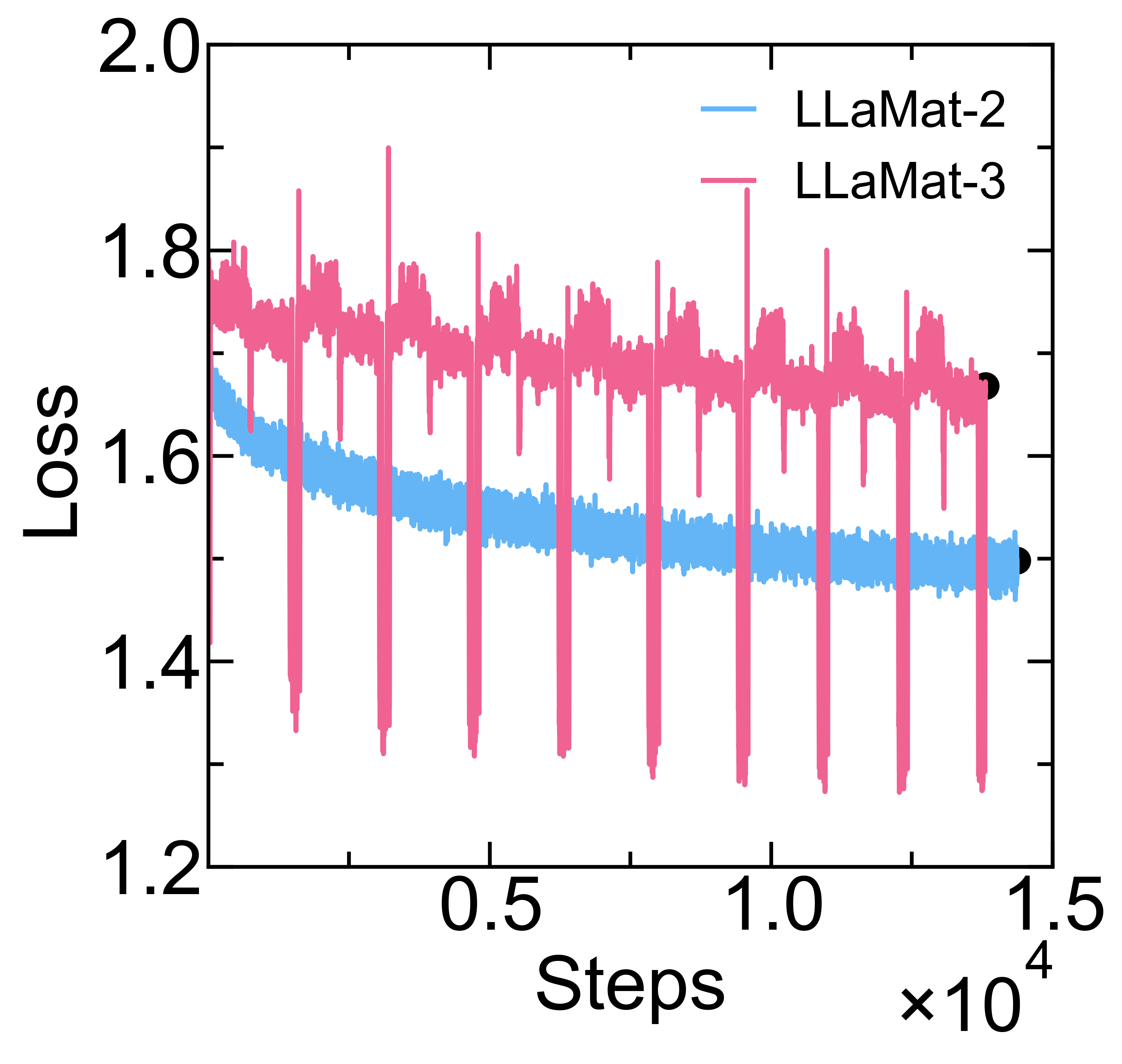}
    \caption{Loss Curve for pretraining}
    \label{fig:loss_llamat_pretrain}
\end{figure}

The pretraining to obtain \llamattwo{} and \llamatthree{} models was performed for 14369 and 13812 steps, respectively. The details of learning rates, warmup ratio, epochs, and the learning rate scheduler are mentioned in Table \ref{tab:hyperparameter_cpt}. Considering the stability of the \llamattwo{} model from the loss curve shown in Fig. \ref{fig:loss_llamat_pretrain}, we took the last checkpoint for further evaluation. In the case of \llamatthree{},  we evaluated intermediate checkpoints to arrive at the final model for downstream evaluation and development of the chat model. The results in table \ref{tab:steps_pretraining_exp} calculated for \llamathree{} were computed just after CPT and before any instruction-finetuning for chat capabilities was done. This experiment informed that the last checkpoint of \llamathree{}, i.e., after 13812 steps, is the best one, and hence, we chose it as our base \llamathree{} model. 

\begin{table}[h!]
\centering
\caption{Hyperparameter details for pretraining of LLAMaT-2 and LLAMaT-3}
\begin{tabular}{ccc}
\hline
\textbf{Hyperparameters}         & \textbf{\llamattwo{}} & \textbf{\llamathree{}} \\ \hline
max\_lr  & $3\text{e-04}$    & $7\text{e-05}$   \\
warmup\_ratio & 0.1            & $0.1$          \\  
min\_lr  & $3\text{e-05}$    & $7\text{e-06}$   \\
epoch    & $1$              & $1$              \\
scheduler & cosine          & cosine            \\ \hline
\end{tabular}
\label{tab:hyperparameter_cpt}
\end{table}


\begin{table}[H]
\centering
\caption{Results on downstream dataset after direct finetuning of the pretrained models}
\scalebox{0.9}{%
    \begin{tabular}{ccccc}
    \toprule
\textbf{Model} & \textbf{MatNLP-Micro-F1} & \textbf{MatNLP-Macro-F1} & \textbf{English-Micro-F1} & \textbf{English-Macro-F1} \\ \midrule
    4k & 89.035 & 82.57 & 84.54 & 79.93 \\
    8k & 88.731 & 82.91 & 83.015 & 78.38 \\
    13k & 89.595 & 84.349 & 84.707 & 80.282 \\
    13812 & 90.02 & 84.752 & 84.06 & 79.547 \\

    \bottomrule
    \end{tabular}}
    
    \label{tab:steps_pretraining_exp}
\end{table}

\subsection{Finetuning}\label{app:finetune_los}

This section shows the loss curves obtained after CIF-IFT of \llamat\ on the CIF-IFT dataset. It can be seen in Figure \ref{fig:cif_loss} a and b that the minimum validation loss occurred at 17000 and 15000 steps, respectively. These models were further used to perform the parameter efficient finetuning to evaluate the performance of the crystal generator on the unconditional crystal structure generation task \cite{gruver2024fine}.

\begin{figure}[H]
    \centering
    \includegraphics[width=0.8\linewidth]{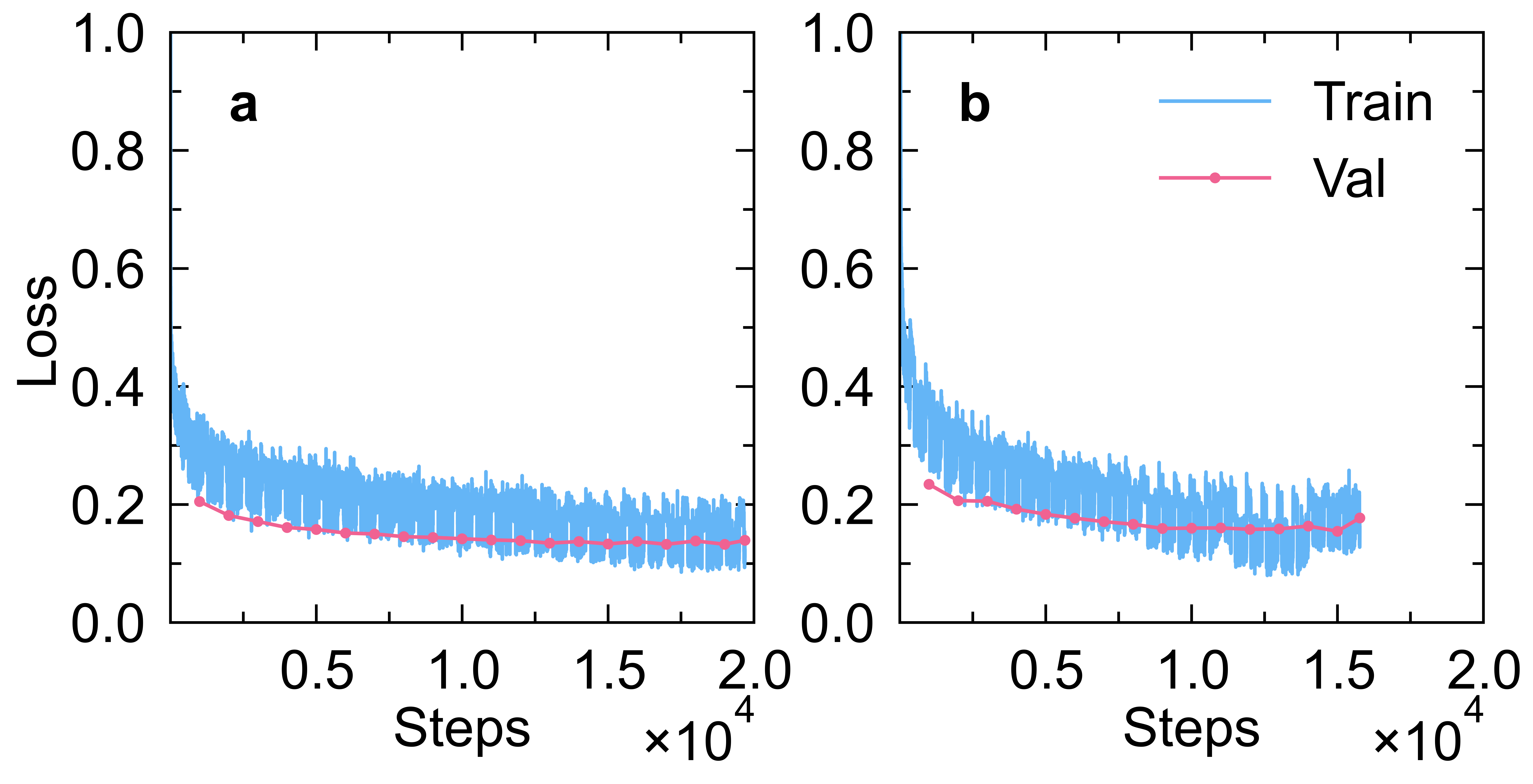}
    \caption{Visualizing the loss curves of (a) \llamat-2-CIF and (b) \llamat-3-CIF models}
    \label{fig:cif_loss}
\end{figure}

\newpage
\subsection{Hardware setup and training time}\label{appendix:hardware}

The training times and hardware setup for each task are as follows :

\begin{itemize}
    \item Pretraining \textsc{LLaMat}-2: 8 NVIDIA A100 80GB GPUs for \textasciitilde 17 days
    \item Pretraining \textsc{LLaMat}-3: 2 CS2 Cerebras Wafer Scale Cluster for \textasciitilde 3 days
    \item \llamat-IE-Copilot (see \ref{method:ift})
    \begin{enumerate}
    \item Instruction fine tuning (stage 1): \textasciitilde8 hours on 8 NVIDIA-A100 80GB GPUs. 
    \item Instruction fine tuning (stage 2): \textasciitilde 1 hour 30 minutes on 8 NVIDIA-A100 80GB GPUs.
    \item Task finetuning (stage 3): 1 hour 10 minutes on NVIDIA-A100 80GB GPUs.
    \end{enumerate}
    \item \llamat-CIF: 2 CS2 Cerebras Wafer Scale Cluster for \textasciitilde3 days
\end{itemize}

For continuous pretraining of \llama-2\ models, we have used 8 NVIDIA A100 80GB GPUs as mentioned above. Since the dataset size and number of parameters are quite large, we use a distributed training methodology to efficiently utilize the storage and computational resources. Table \ref{tab:pptpdp} lists our experiments to obtain optimal levels of data (DP), tensor (TP), and pipeline parallelisms (PP). We achieved the best token consumption rate of ~27.1k tokens/second by considering PP=4, TP=1, and DP=2. Based on these experiments, we can also state that TP was less effective in our case than DP and PP.  

In the case of \llamathree{} pretraining, we used 2 CS2 Cerebras Wafer Scale Cluster. Here, we did not require parallelism as used in GPUs because of the linear scaling of the compute performance with change in the number of accelerators\cite{cerebras_gpu_linear}. During the pretraining, we used batch\_size of 960 and micro\_batch\_size of 80 as suggested by the training script provided at \href{https://github.com/Cerebras/modelzoo}{Cerebras Model Zoo on GitHub}.

\begin{table}[h!]
\centering
\begin{tabular}{cccccc}
\toprule
\textbf{Nodes x GPUs} & \textbf{DP} & \textbf{TP} & \textbf{PP} & \textbf{tokens/s (k)} & \textbf{tokens/s/gpu (k)} \\ 
\midrule
1x2 & 1 & 1 & 2 & 7.5 & 3.75 \\ 
1x2 & 1 & 2 & 1 & 4.8 & 2.4 \\ 
1x2 & 2 & 1 & 1 & OOM & OOM \\ 
1x4 & 1 & 1 & 4 & 12 & 3 \\ 
1x4 & 1 & 1 & 1 & 5.6 & 1.4 \\ 
1x4 & 4 & 1 & 1 & OOM & OOM \\ 
1x4 & 2 & 2 & 1 & 9.4 & 2.35 \\ 
1x4 & 2 & 1 & 2 & 13.8 & 3.45 \\ 
1x4 & 1 & 2 & 2 & 12.5 & 3.125 \\ 
1x8 & 1 & 1 & 8 & 22.9 & 2.8625 \\ 
1x8 & 1 & 1 & 1 & 5 & 0.625 \\ 
1x8 & 8 & 1 & 1 & OOM & OOM \\ 
1x8 & 2 & 2 & 2 & 17.5 & 2.1875 \\ 
1x8 & 1 & 2 & 4 & 14.2 & 1.775 \\ 
1x8 & 1 & 1 & 4 & 9.9 & 1.2375 \\ 
1x8 & 2 & 4 & 2 & 23.4 & 2.925 \\ 
1x8 & 1 & 2 & 4 & 13 & 1.625 \\ 
1x8 & 2 & 4 & 1 & 14.9 & 1.8625 \\ 
1x8 & 4 & 2 & 1 & 21.4 & 2.675 \\ 
\textbf{1x8} & \textbf{2} & \textbf{4} & \textbf{1} & \textbf{27.1} & \textbf{3.3875} \\
\bottomrule
\end{tabular}
\caption{GPU Performance Metrics. OOM stands for Out-Of Memory error.}
\label{tab:pptpdp}
\end{table}

\newpage
\section{Dataset distribution optimization}\label{appendix:section:dataset_optimization}
\vspace{-0.3in}
\subsection{Pretraining}
\vspace{-0.7in}
\begin{table}[H]
    \centering
    \caption{Details of pretraining datasets for obtaining \llamattwo{} and \llamatthree{}}
    \begin{tabular}{ccccccc}
    \toprule
         &  \multicolumn{2}{c}{\# samples}&  \multicolumn{2}{c}{\# tokens (\llamatwo)}&  \multicolumn{2}{c}{\# tokens (\llamathree)}\\
         \midrule
         Dataset&  train&  val&  train&  val&  train& val\\
         \midrule
         P1&   2,686,786&  &   18,872,303,847 &   &  & \\
         P2&  1,055,330&  106,395&  9,050,611,308 &  413,927,438&   7,831,900,364&  442,507,226\\
         P3&  225,634&  &  1,864,471,418&  &  1,589,414,318& -\\
         \midrule
         MSCD&  36,875&  &  5,975,502&  0&  5,212,659& 0\\
         \midrule
         RedPajama&  651,356&  279,158&  962,319,047&  414,815,173&  805,636,840& 347,375,685\\
         \midrule
         CIF&  470,222 &  9,598&  788,427,184&  16,124,004&  633,237,003& 12,947,445\\
         \midrule
         \textbf{Total} &  \textbf{5,126,203} &  \textbf{395,151} &  \textbf{31,544,108,306} &  \textbf{844,866,615} &  \textbf{10,865,401,184} & \textbf{802,830,356} \\
         \bottomrule
    \end{tabular}
    \label{tab:cpt_data}
\end{table}

\subsection{Finetuning}
The first step in instruction finetuning our models is training on OpenOrca, a general instruction finetuning dataset. We trained the model for different steps between 0-800k, then finetuned it on the downstream dataset again before evaluation.

Table \ref{tab:orca_llamat2} shows the results on \llamat-3 and and \llamat-2. We observed that \llamat-2's English capability increases with more steps in general, while for \llamat-3, there is no such observation. Also, \llamat-3's score in MatNLP is lower than its score at 0 steps. This could be because OpenOrca is a general-purpose IFT dataset unrelated to our downstream tasks. Since \llamat-3 already had a high score on both English and MatNLP initially, we don't notice a significant further increase. From the results of \llamat-3\ \ref{tab:orca_llamat2}, we decided to fix 576k training samples for Open-Orca instruction finetuning for \llamat-3, and 448k training samples for \llamat-2. Further IFT processes are described in the methodology section.

\begin{table}[h]
\centering
 \caption{Performance of \llamat-2 and \llamat-3 on MatNLP and English validation sets after instruction-finetuning on Open-Orca dataset to varying degrees. The optimal dataset size is chosen based on the Pareto optimal performance on both MatNLP and Eng datasets.}
    \label{tab:orca_llamat2}
\scalebox{1}{%
    \begin{tabular}{cccccc}
    \toprule
\textbf{Steps} & \textbf{MicroF1-MatNLP} & \textbf{MacroF1-MatNLP} & \textbf{MicroF1-Eng} & \textbf{MacroF1-Eng} \\ 
    \midrule
    \textbf{\llamat-2} & & & & \\
    \midrule
       0k & 87.85 & 82.26 & 82.23 & 78.73 \\ 
      64k & 88.44 & 83.07 & 82.94 & 79.32 \\ 
     128k & 88.72 & 83.35 & 83.31 & 79.47 \\ 
     192k & 89.08 & 83.71 & 83.20 & 79.55 \\ 
     256k & 89.34 & 84.09 & 83.60 & 79.79 \\ 
     320k & 88.14 & 82.68 & 84.22 & 80.32 \\ 
     384k & 88.48 & 83.54 & 84.05 & 80.43 \\ 
     448k & 89.51 & 84.66 & 83.69 & 80.04 \\ 
     512k & 89.07 & 83.96 & 84.04 & 80.30 \\ 
     576k & 89.09 & 83.76 & 84.47 & 80.82 \\ 
     640k & 88.60 & 83.30 & 84.95 & 81.30 \\ 
     768k & 89.23 & 84.05 & 84.34 & 80.55 \\ 
     800k & 88.48 & 83.12 & 85.02 & 81.22 \\ 
    \midrule
    \textbf{\llamat-3} & & & & \\
    \midrule
      0k & 89.70 & 83.71 & 84.56 & 80.24 \\ 
     64k & 88.40 & 82.85 & 85.31 & 80.57 \\ 
    128k & 86.39 & 80.29 & 83.63 & 79.24 \\ 
    192k & 88.48 & 82.67 & 84.20 & 79.38 \\ 
    256k & 85.97 & 80.32 & 84.68 & 79.81 \\ 
    320k & 88.03 & 82.10 & 85.10 & 80.49 \\ 
    384k & 87.42 & 81.95 & 85.40 & 80.67 \\ 
    448k & 86.85 & 81.64 & 85.06 & 80.25 \\ 
    512k & 87.89 & 82.37 & 84.74 & 80.24 \\ 
    576k & 88.79 & 83.09 & 84.74 & 80.01 \\ 
    640k & 88.40 & 82.85 & 85.31 & 80.57 \\ 
    768k & 86.96 & 81.27 & 85.50 & 80.63 \\ 
    800k & 87.70 & 82.48 & 85.07 & 80.19 \\ 
    \bottomrule
    \end{tabular}}
   
\end{table}

We also conducted experiments with different training samples for the MathQA and honeybee datasets for \llamattwo{}.

\begin{table}[H]
\centering
\caption{Results for training with MathQA and different sample size of honeybee dataset on downstream evaluation}
\scalebox{0.7}{%
    \begin{tabular}{ccccccccc}
    \toprule
\textbf{Model} & \textbf{Pretrain} & \textbf{OpenOrca} & \textbf{MathQA} & \textbf{Honeybee} & \textbf{MicroF1-MatNLP} & \textbf{MacroF1-MatNLP} & \textbf{MicroF1-English} & \textbf{MacroF1-English} \\ \midrule
    \llama-2 & & & & & 84.24 & 77.75 & 80.63 & 77.01 \\ 
    \llamat & 10B & 0 & 0 & 0 & 85.43 & 79.68 & 78.8 & 75.33 \\ 
    \llamat & 30B & 0 & 0 & 0 & 87.85 & 82.26 & 82.23 & 78.73 \\ 
    \midrule
    \llamat & 30B & 448k & 0 & 0 & 89.51 & 84.66 & 83.69 & 80.04 \\ 
    \llamat & 30B & 448k & 0 & 32k & 88.52 & 83.24 & 83.25 & 79.48 \\ 
    \llamat & 30B & 448k & 0 & 48k & 88.52 & 83.02 & 84.5 & 80.8 \\ 
    \llamat & 30B & 448k & 0 & 96k & 88.6 & 83.04 & 83.97 & 80.17 \\ 
    \llamat & 30B & 448k & 0 & 144k & 88.44 & 83.12 & 84.38 & 80.5 \\ 
    \midrule
    \llamat & 30B & 448k & 7500*3 & 0  & 89.66 & 84.77 & 82.59 & 78.67\\ 
    \llamat & 30B & 448k & 7500*3 & 32k & 87.89 & 82.27 & 83.56 & 79.66 \\ 
    \llamat & 30B & 448k & 7500*3 & 48k & 88.28 & 82.9 & 84.37 & 80.68 \\ 
    \llamat & 30B & 448k & 7500*3 & 96k & 88.04 & 82.39 & 84.17 & 80.25 \\ 
    \llamat & 30B & 448k & 7500*3 & 144k & 88.24 & 82.8 & 83.8 & 79.84 \\     
    \bottomrule
    \end{tabular}}
    \label{tab:hbmx}
\end{table}


\section{Model Performance}
\label{appendix:section:all_results}

\begin{table}[H]
\centering
 \caption{F1-score results on all our datasets. SIE = Structured information extraction. FT = Finetuned}
\scalebox{0.6}{
\renewcommand{\arraystretch}{1.1} 
 \begin{tabular}{cccccccccc}
 \toprule
\textbf{Task} & \textbf{sub-dataset} & \textbf{\shortstack{\llamat-3\\chat}} & \textbf{\llamat-3} & \textbf{\shortstack{\llama-3\\chat FT}} & \textbf{\shortstack{\llama-3\\FT}} & \textbf{\shortstack{\llamat-2\\chat}} & \textbf{\llamat-2} & \textbf{\shortstack{\llamat-2\\chat FT}} & \textbf{\shortstack{\llama-2\\FT}} \\ \midrule
\textbf{MatNLP} & \textbf{Micro-F1} & & & & & & & &  \\
\midrule

    EntityRecognition & Matscholar    & 88.68 & 84.72 & 84.24 & 85.09 & 85.22 & 83.62 & 83.12 & 82.43 \\ 
    EntityRecognition & SOFC-1        & 91.62 & 88.99 & 88.5 & 90.95 & 90.63 & 87.41 & 88.9 & 88.51 \\ 
    EntityRecognition & SOFC-2        & 86.05 & 81.58 & 81.11 & 84.11 & 85.95 & 85.71 & 82.5 & 80.55 \\ 
    EntityRecognition & SC-CoMIcs-1   & 90.48 & 87.74 & 79.58 & 90 & 91.98 & 91.61 & 90.82 & 91.1 \\ 
    Classification & Glass            & 94.33 & 84.33 & 86.67 & 92 & 93.33 & 92.33 & 92 & 92 \\ 
    Classification & SynthesisActions & 96.44 & 96.13 & 95.76 & 96.22 & 96.68 & 95.76 & 96.96 & 96.22 \\ 
    Classification & SOFC-3           & 94.18   & 93.05 & 92.62 & 93.8 & 93.75 & 94.07 & 93.69 & 93.64 \\ 
    EntityExtraction & SC-CoMIcs-2    & 94.47 & 80.93 & 83.2 & 94.25 & 95.96 & 95.33 & 95.2 & 92.17 \\ 
    EntityExtraction & SC-CoMIcs-3    & 94.02 & 93.54 & 70.57 & 74.64 & 99.76 & 99.76 & 99.76 & 100 \\ 
    EntityExtraction & MatSci         & 100 & 100 & 100 & 100 & 100 & 100 & 100 & 100 \\ 
    All MatNLP & Mean Micro-F1 &  93.0270 & 89.101 & 86.2250 & 90.106 & 93.3260 & 92.5600 & 92.2950 & 91.662 \\
     \midrule
    \textbf{MatNLP} & \textbf{Macro-F1} & & & & & & & &  \\
    \midrule
    EntityRecognition & Matscholar     & 85.33 & 79.96 & 78.66 & 78.87 & 80.22 & 79.85 & 79.87 & 77.42 \\ 
    EntityRecognition & SOFC-1         & 80.07 & 75.76 & 76.1 & 78.61 & 79.5 & 77.11 & 78.0 & 77.94 \\ 
    EntityRecognition & SOFC-2         & 77.62 & 73.3 & 72.49 & 73.22 & 78.94 & 76.41 & 71.04 & 70.5 \\ 
    EntityRecognition & SC-CoMIcs-1    & 87.91 & 85.07 & 76.64 & 87.06 & 89.26 & 88.58 & 88.13 & 88.44 \\ 
    Classification & Glass             & 93.36 & 83.04 & 85.28 & 90.65 & 92.16 & 90.72 & 90.22 & 90.22 \\ 
    Classification & SynthesisActions  & 94.76 & 94.3 & 93.4 & 93.74 & 94.77 & 94.01 & 95.96 & 94.3 \\ 
    Classification & SOFC-3            & 80.54 & 78.37 & 79.08 & 78.96 & 78.17 & 77.09 & 77.91 & 73.72 \\ 
    EntityExtraction & SC-CoMIcs-2     & 92.07 & 73.18 & 76.46 & 91.07 & 94.52 & 93.61 & 93.04 & 88.6 \\ 
    EntityExtraction & SC-CoMIcs-3     & 93.89 & 92.5 & 49.64 & 66.36 & 99.81 & 99.81 & 99.81 & 100.0 \\ 
    EntityExtraction & MatSci          & 100.0 & 100.0 & 100.0 & 100.0 & 100.0 & 100.0 & 100.0 & 100.0 \\ 
     All MatNLP & Mean Macro-F1        & 88.46 & 83.55 & 78.78 & 83.85 & 88.74 & 87.72 & 87.40 & 86.11 \\ 
     \midrule
\textbf{English} & \textbf{Micro-F1} & & & & & & & &  \\
\midrule
    QnA &  SQuAD        & 85.26 & 84.71 & 85.11 & 85.05 & 86.46 & 85.68 & 85.02 & 86.03  \\ 
    MCQ &  HellaSwag    & 78.18 & 83.5 & 84.22 & 84.22 & 81.86 & 81.66 & 82.28 & 83.2  \\ 
    MCQ &  BoolQ        & 84.2 & 84.8 & 86.0 & 86.0 & 85.6 & 85.4 & 84.4 & 86.2  \\ 
    MCQ &  Story-Cloze  & 98.46 & 98.15 & 98.76 & 97.53 & 97.53 & 96.3 & 97.53 & 96.91  \\ 
    All English &  Mean Micro-F1 & 86.52 & 87.79 & 88.52 & 88.2 & 87.86 & 87.26 & 87.31 & 88.09  \\ 
    \midrule
\textbf{English} & \textbf{Macro-F1} & & & & & & & &  \\
\midrule
    QnA &  SQuAD        & 72.52 & 72.06 & 72.9 & 72.34 & 74.1 & 73.82 & 72.52 & 74.56 \\ 
    MCQ &  HellaSwag    & 78.26 & 83.45 & 84.08 & 84.17 & 81.7 & 81.42 & 82.12 & 83.02 \\ 
    MCQ &  BoolQ        & 83.75 & 83.8 & 85.02 & 85.26 & 85.16 & 84.75 & 83.61 & 85.37 \\ 
    MCQ &  Story-Cloze  & 98.46 & 98.15 & 98.76 & 97.53 & 97.53 & 96.3 & 97.53 & 96.91 \\ 
    All English & Mean Macro-F1 & 83.25 & 84.36	& 85.19	& 84.82 & 84.62	& 84.072 & 83.94 & 84.96 \\
    \midrule
 \textbf{SIE Doping} & \textbf{F1} & & & & & & & &  \\
 \midrule
 
    NER & basemats & 0.818 & 0.901 & 0.865 & 0.8 & 0.836 & 0.819 & 0.843 & 0.859 \\ 
    NER & dopants & 0.908 & 0.857 & 0.87 & 0.91 & 0.859 & 0.833 & 0.823 & 0.833 \\ 
    RE & triplets & 0.782 & 0.814 & 0.749 & 0.777 & 0.763 & 0.764 & 0.751 & 0.73 \\ 
    All & exact-match & 0.619 & 0.587 & 0.571 & 0.571 & 0.571 & 0.524 & 0.508 & 0.603 \\ 
     \midrule
 \textbf{SIE General} & \textbf{F1} & & & & & & & &  \\
 \midrule
    NER & acronym & 0.353 & 0.353 & 0.111 & 0.154 & 0.133 & 0 & 0 & 0 \\ 
    NER & applications & 0.571 & 0.621 & 0.471 & 0.596 & 0.682 & 0.696 & 0.671 & 0.634 \\ 
    NER & name & 0.338 & 0.347 & 0.417 & 0.406 & 0.32 & 0.212 & 0.31 & 0.328 \\ 
    NER & formula & 0.6 & 0.511 & 0.604 & 0.629 & 0.661 & 0.716 & 0.631 & 0.679 \\ 
    NER & structure or phase & 0.403 & 0.484 & 0.169 & 0.305 & 0.693 & 0.728 & 0.525 & 0.526 \\ 
    NER & description & 0.365 & 0.375 & 0.261 & 0.357 & 0.393 & 0.385 & 0.34 & 0.343 \\ 
    RE & formula-name & 0 & 0.222 & 0 & 0.235 & 0.125 & 0.125 & 0.095 & 0.1 \\ 
    RE & formula-structure/phase & 0.167 & 0.275 & 0.121 & 0.187 & 0.567 & 0.609 & 0.371 & 0.34 \\ 
    RE & formula-application & 0.435 & 0.579 & 0.413 & 0.631 & 0.574 & 0.6 & 0.61 & 0.556 \\ 
    RE & formula-description & 0.182 & 0.306 & 0.118 & 0.27 & 0.255 & 0.385 & 0.245 & 0.242 \\ 
     \midrule
 \textbf{SIE MOFs} & \textbf{F1} & & & & & & & &  \\
 \midrule
    NER & name of mof & 0.667 & 0.683 & 0.7 & 0.713 & 0.736 & 0.742 & 0.742 & 0.812 \\ 
    NER & mof formula & 0.462 & 0.313 & 0.626 & 0.611 & 0.646 & 0.66 & 0.707 & 0.733 \\ 
    NER & mof description & 0.337 & 0.388 & 0.398 & 0.447 & 0.466 & 0.503 & 0.358 & 0.422 \\ 
    NER & guest species & 0.364 & 0.323 & 0.421 & 0.571 & 0.783 & 0.809 & 0.514 & 0.471 \\ 
    NER & applications & 0.654 & 0.638 & 0.665 & 0.638 & 0.674 & 0.679 & 0.627 & 0.665 \\ 
    NER & exact-match & 0.098 & 0.078 & 0.118 & 0.118 & 0.118 & 0.098 & 0.078 & 0.118 \\ 
    RE & name-formula & 0 & 0 & 0 & 0 & 0 & 0 & 0 & 0 \\ 
    RE & name-guestspecies & 0.195 & 0.162 & 0.286 & 0.292 & 0.667 & 0.621 & 0.298 & 0.261 \\ 
    RE & name-application & 0.318 & 0.424 & 0.425 & 0.383 & 0.407 & 0.461 & 0.401 & 0.495 \\ 
    RE & name-description & 0.204 & 0.324 & 0.4 & 0.286 & 0.321 & 0.295 & 0.392 & 0.302 \\ 
    \midrule
    \textbf{DiSCoMaT} & \textbf{Accuracy} & & & & & & & &  \\
    \midrule
    table & comptable & 0.846 & 0.825 & 0.566 & 0.835 & 0.87 & 0.837 & 0.828 & 0.828 \\ 
    table & regex & 0.836 & 0.867 & 0.195 & 0.824 & 0.878 & 0.856 & 0.844 & 0.848 \\ 
    table & gid & 0.772 & 0.795 & 0.867 & 0.802 & 0.872 & 0.847 & 0.78 & 0.809 \\ 
    table & composition & 0.245 & 0.345 & 0.545 & 0.359 & 0.595 & 0.596 & 0.629 & 0.631 \\ 
    table & chemical & 0.508 & 0.647 & 0.694 & 0.587 & 0.704 & 0.678 & 0.659 & 0.661 \\ 
    All & exact-match & 405/602 & 397/578 & 180/375 & 411/610 & 547/728 & 534/727 & 541/727 & 538/728 \\ 
 \bottomrule
 \end{tabular}}
 \label{appendix:all_results}
\end{table}

\begin{table}
\centering
\caption{F1-score results comparison for our models and some closed-source models}
\scalebox{0.6}{%
    \begin{tabular}{ccccccccccc}
    \toprule
\textbf{sub-dataset} & \textbf{\shortstack{\llamat-3\\chat}} & \textbf{\shortstack{\llamat-2\\chat}} & \textbf{\shortstack{Claude-3.5-\\Sonnet}} & \textbf{\shortstack{Claude-3-\\Opus}} & \textbf{\shortstack{Claude-3-\\Haiku}} & \textbf{\shortstack{Gemini-1.5\\-Pro}} & \textbf{\shortstack{Gemini-1.5-\\Flash}} & \textbf{\shortstack{Gemini-1.5-\\Flash-8b}} & \textbf{\shortstack{GPT-4o}} & \textbf{\shortstack{GPT-4}} \\ \midrule
    \textbf{MatNLP} & \textbf{Micro-F1} & & & & & & & &  \\
\midrule
    Matscholar & 88.68 & 85.22 & 37.92 & 35.11 & 0.21 & 39.5 & 10.22 & 17.34 & 20.46 & 24.06 \\ 
    SOFC-1 & 91.62 & 90.63 & 81.74 & 71.3 & 5.72 & 81.07 & 1.78 & 50.4 & 80.76 & 78.49 \\ 
    SOFC-2 & 86.05 & 85.95 & 70.75 & 63.47 & 2.45 & 12.2 & 0 & 3.39 & 76.22 & 74.73 \\ 
    SC-CoMIcs-1 & 90.48 & 91.98 & 55.29 & 50.6 & 21.81 & 51.84 & 53.6 & 48.37 & 45.79 & 48.11 \\ 
    Glass & 94.33 & 93.33 & 60.33 & 55.17 & 55.33 & 68.33 & 72.67 & 77.67 & 74 & 66 \\ 
    SynthesisActions & 96.44 & 96.68 & 75.07 & 64.44 & 20.44 & 71.65 & 65.8 & 57.57 & 68.23 & 57.88 \\ 
    SOFC-3 & 94.18 & 93.75 & 39.25 & 23.75 & 21.35 & 27.82 & 45.23 & 52.78 & 70.4 & 51.32 \\ 
    SC-CoMIcs-2 & 94.47 & 95.96 & 92.71 & 92.62 & 81.64 & 91.92 & 84.12 & 84.56 & 95.91 & 82.28 \\ 
    SC-CoMIcs-3 & 94.02 & 99.76 & 52.87 & 20.99 & 56.22 & 53.35 & 33.97 & 20.81 & 52.39 & 18.18 \\ 
    MatSci & 100 & 100 & 100 & 99.85 & 93.46 & 99.54 & 99.74 & 98.63 & 98.3 & 99.35 \\ 
    Mean Micro-F1 & 93.03 & 93.33 & 66.59 & 57.73 & 35.86 & 59.72 & 46.71 & 51.15 & 68.25 & 60.04 \\ 

         \midrule
             \textbf{MatNLP} & \textbf{Macro-F1} & & & & & & & &  \\
\midrule
   Matscholar & 85.33 & 80.22 & 33.38 & 21.85 & 0.06 & 29.18 & 6.04 & 6.96 & 12.78 & 16.01 \\ 
    SOFC-1 & 80.07 & 79.5 & 71.44 & 57.9 & 5.66 & 66.42 & 1.37 & 36.09 & 69.98 & 63.69 \\ 
    SOFC-2 & 77.62 & 78.94 & 63.64 & 62.42 & 2.28 & 10.63 & 0.0 & 2.52 & 71.14 & 61.56 \\ 
    SC-CoMIcs-1 & 87.91 & 89.26 & 48.52 & 46.01 & 19.9 & 51.45 & 46.91 & 42.36 & 41.51 & 41.62 \\ 
    Glass & 93.36 & 92.16 & 60.32 & 55.13 & 55.26 & 67.94 & 72.01 & 76.7 & 73.32 & 65.7 \\ 
    SynthesisActions & 94.76 & 94.77 & 65.74 & 54.21 & 20.97 & 61.42 & 61.38 & 51.38 & 59.4 & 53.73 \\ 
    SOFC-3 & 80.54 & 78.17 & 36.22 & 23.51 & 21.25 & 27.05 & 40.46 & 45.62 & 55.48 & 44.65 \\ 
    SC-CoMIcs-2 & 92.07 & 94.52 & 89.42 & 90.25 & 74.23 & 87.76 & 76.77 & 74.53 & 94.48 & 75.4 \\ 
    SC-CoMIcs-3 & 92.89 & 99.81 & 52.88 & 20.43 & 44.27 & 40.41 & 35.41 & 19.41 & 50.58 & 18.88 \\ 
    MatSci & 100.0 & 100.0 & 100.0 & 99.74 & 85.91 & 98.11 & 99.55 & 97.57 & 95.53 & 98.86 \\
    Mean Macro-F1 & 88.46 & 88.73 & 62.16 & 53.145 & 32.98 & 54.037 & 43.99 & 45.31 & 62.42 & 54.01\\ 

     \midrule
 \textbf{SIE Doping} & \textbf{F1} & & & & & & & &  \\
 \midrule
    basemats & 0.818 & 0.836 & 0.701 & 0.716 & 0.691 & 0.707 & 0.773 & 0.663 & 0.685 & 0.61 \\ 
    dopants & 0.908 & 0.859 & 0.743 & 0.751 & 0.753 & 0.739 & 0.733 & 0.795 & 0.798 & 0.78 \\ 
    triplets & 0.782 & 0.763 & 0.591 & 0.601 & 0.586 & 0.597 & 0.609 & 0.615 & 0.594 & 0.609 \\ 
    exact-match & 0.619 & 0.571 & 0.311 & 0.371 & 0.138 & 0.397 & 0.288 & 0.123 & 0.371 & 0.148 \\ 
         \midrule
 \textbf{SIE General} & \textbf{F1} & & & & & & & &  \\
 \midrule
    acronym & 0.353 & 0.133 & 0.24 & 0.143 & 0.12 & 0.235 & 0.19 & 0.1 & 0.273 & 0.176 \\ 
    applications & 0.571 & 0.682 & 0.236 & 0.095 & 0.124 & 0.182 & 0.335 & 0.387 & 0.253 & 0.135 \\ 
    name & 0.338 & 0.32 & 0.111 & 0.071 & 0.034 & 0.048 & 0.229 & 0.055 & 0.032 & 0.065 \\ 
    formula & 0.6 & 0.661 & 0.239 & 0.386 & 0.238 & 0.305 & 0.302 & 0.374 & 0.316 & 0.332 \\ 
    structure or phase & 0.403 & 0.693 & 0.137 & 0.174 & 0.101 & 0.075 & 0.161 & 0.193 & 0.236 & 0.075 \\ 
    description & 0.365 & 0.393 & 0.035 & 0.041 & 0.044 & 0.031 & 0.045 & 0.026 & 0.037 & 0.03 \\ 
    formula-name & 0 & 0.125 & 0.059 & 0.038 & 0.062 & 0 & 0.045 & 0.017 & 0 & 0.036 \\ 
    formula-structure/phase & 0.167 & 0.567 & 0.039 & 0.036 & 0.031 & 0.04 & 0.041 & 0.071 & 0.033 & 0.017 \\ 
    formula-application & 0.435 & 0.574 & 0.121 & 0.024 & 0.067 & 0.034 & 0.19 & 0.199 & 0.092 & 0.066 \\ 
    formula-description & 0.182 & 0.255 & 0.007 & 0.006 & 0.013 & 0.008 & 0.013 & 0.014 & 0.011 & 0.01 \\ 
         \midrule
 \textbf{SIE MOFs} & \textbf{F1} & & & & & & & &  \\
 \midrule
    name of mof & 0.667 & 0.736 & 0.525 & 0.431 & 0.645 & 0.318 & 0.582 & 0.248 & 0.541 & 0.483 \\ 
    mof formula & 0.462 & 0.646 & 0.229 & 0.447 & 0.278 & 0.405 & 0.203 & 0.26 & 0.149 & 0.182 \\ 
    mof description & 0.337 & 0.466 & 0.047 & 0.064 & 0.034 & 0.05 & 0.028 & 0.022 & 0.042 & 0.04 \\ 
    guest species & 0.364 & 0.783 & 0.372 & 0.361 & 0.306 & 0.451 & 0.336 & 0.469 & 0.48 & 0.28 \\ 
    applications & 0.654 & 0.674 & 0.441 & 0.393 & 0.431 & 0.387 & 0.366 & 0.356 & 0.421 & 0.452 \\ 
    exact-match & 0.098 & 0.118 & 0 & 0 & 0 & 0 & 0 & 0 & 0 & 0 \\ 
    name-formula & 0 & 0 & 0.1 & 0.043 & 0.348 & 0.308 & 0.083 & 0.071 & 0.211 & 0.04 \\ 
    name-guestspecies & 0.195 & 0.667 & 0.273 & 0.276 & 0.179 & 0.343 & 0.215 & 0.291 & 0.293 & 0.181 \\ 
    name-application & 0.318 & 0.407 & 0.16 & 0.111 & 0.094 & 0.128 & 0.138 & 0.06 & 0.215 & 0.163 \\ 
    name-description & 0.204 & 0.321 & 0.016 & 0.009 & 0.011 & 0.019 & 0.005 & 0.001 & 0.012 & 0.008 \\ 
        \midrule
    \textbf{DiSCoMaT} & \textbf{Accuracy} & & & & & & & &  \\
    \midrule
    comptable & 0.846 & 0.87 & 0.852 & 0.817 & 0.748 & 0.846 & 0.773 & 0.803 & 0.796 & 0.806 \\ 
    regex & 0.836 & 0.878 & 0.4 & 0.403 & 0.311 & 0.439 & 0.326 & 0.436 & 0.414 & 0.466 \\ 
    gid & 0.772 & 0.872 & 0.957 & 0.893 & 0.899 & 0.948 & 0.958 & 0.944 & 0.696 & 0.968 \\ 
    composition & 0.245 & 0.595 & 0.814 & 0.849 & 0.53 & 0.827 & 0.688 & 0.548 & 0.723 & 0.867 \\ 
    chemical & 0.508 & 0.704 & 0.723 & 0.804 & 0.498 & 0.755 & 0.781 & 0.654 & 0.602 & 0.668 \\ 
    exact-match & 405/602 & 547/728 & 65/734 & 42/737 & 0/732 & 146/737 & 71/728 & 46/729 & 52/737 & 20/736 \\ 

    \bottomrule
    \end{tabular}}
    \label{app:tab:all_results_external}
\end{table}

\section{DiSCoMat instruction and JSON Schema}
\label{appendix:discomat}
\label{appendix:nerre}
We give the following instructions to the model before providing the question and table from which to answer. It includes the JSON schema of the output format in the form of a dictionary containing non-empty lists. The definition for each entry of the dictionary is also passed to the model. \\
\textbf{Prompt:}
\begin{lstlisting}
    
You are an expert in materials science and extracting data from tables. You have the fill the 
following dictionary for the given table. Each key is defined as follows:
'comp_table'- If the input table has material compositions then return [1], else [0];
'regex_table'- If the input table has material compositions and they can be extracted using a regular expression parser, then return [1], else [0].
'composition_row_index'-The list containing the index of rows which have complete information about material composition.
'chemical_col_index'-The list containing the index of columns which report values of constituent chemicals of the material.
'composition_col_index'-The list containing the index of columns which have complete information about material composition.
'chemical_row_index'-The list containing the index of rows which report values of constituent chemicals of the material.
'gid_row_index'-The index of row having material identifier.
'gid_col_index'-The index of column having material identifier.
 
dictionary =  
{'comp_table': [],
'regex_table': [],
'composition_row_index': [],
'composition_col_index': [],
'chemical_row_index': [],
'chemical_col_index': [],
'gid_row_index': [],
'gid_col_index': []}
NOTE:The output will be the dictionary with keys having non-empty lists ONLY.

\end{lstlisting}
 Sometimes, the output cannot be parsed as a JSON, and we don't consider such cases in the evaluation of our models. The total count of outputs that are parsed can be seen in Table \ref{appendix:all_results} in the 'exact-match' row, which states exact matches / total outputs parsed.


\section{Prompts for MatBookQA}\label{appendix:prompts_matbookqa}

\subsection*{Short Prompts}
\begin{itemize}
    \item You are a materials scientist. Use your expertise to generate concise answers to the following questions.
    \item As a materials scientist, provide short, precise answers to these questions.
    \item With your knowledge in materials science, answer the following questions succinctly.
    \item Given your background in materials science, provide brief, expert answers to these queries.
    \item Using your expertise in materials science, generate short answers for the following questions.
    \item Drawing from your experience in materials science, answer these questions with concise and accurate information.
    \item As an expert in materials science, provide quick, accurate answers to these questions.
    \item From your perspective as a materials scientist, generate short and precise answers to the following questions.
    \item Using your knowledge as a materials scientist, answer these questions briefly and accurately.
    \item Leverage your expertise in materials science to provide concise answers to these queries.
\end{itemize}

\subsubsection*{Long Prompts}
\begin{itemize}
    \item You are a materials scientist. Use your expertise in the field to generate detailed and comprehensive answers for the following questions.
    \item As a materials scientist, provide thorough and well-explained answers to these questions.
    \item With your knowledge in materials science, answer the following questions with detailed and extensive information.
    \item Given your background in materials science, provide long and comprehensive answers to these queries.
    \item Using your expertise in materials science, generate detailed and in-depth answers for the following questions.
    \item Drawing from your experience in materials science, answer these questions with elaborate and accurate information.
    \item As an expert in materials science, provide thorough and well-detailed answers to these questions.
    \item From your perspective as a materials scientist, generate long and comprehensive answers to the following questions.
    \item Using your knowledge as a materials scientist, answer these questions in detail and with full explanations.
    \item Leverage your expertise in materials science to provide extensive and well-explained answers to these queries.
\end{itemize}

\section{CIF IFT prompts}\label{appendix:cif_prompts}
\subsection{Syntactic tasks}
\begin{itemize}
    \item You are a Material Science expert who works with crystallographic files (CIF files). Use your understanding of the CIF file format to extract information about the unit cell structure.
    \item Utilize your expertise in Material Science to extract data regarding the unit cell structure from CIF files, drawing upon your comprehension of the file format.
    \item As a specialist in Material Science, employ your knowledge of CIF files to extract pertinent details concerning the unit cell structure.
    \item As a Material Science expert, utilize CIF file parsing to extract essential data regarding the unit cell configuration.
    \item Draw upon your Material Science expertise to extract unit cell structure information from CIF files, utilizing your understanding of the file format.
    \item Employ your understanding of Material Science and CIF file format to extract crucial information concerning the unit cell arrangement.
    \item As a specialist in Material Science, employ CIF file analysis to gather insights into the unit cell structure.
    \item Utilize your proficiency in Material Science to parse CIF files and extract relevant details regarding the unit cell configuration.
    \item Draw upon your expertise in Material Science to extract insights into the unit cell structure by analyzing CIF files.
\end{itemize}

\subsection{Semantic tasks}
\subsubsection{Generative tasks}
\begin{itemize}
    \item You are a Material Science expert who works with crystallographic files (CIF files). Use your expertise to answer the following question related to the generation of stable materials when some information about it is described.
    \item Employ your expertise in Material Science, particularly in working with CIF files, to address the question concerning the creation of stable materials with partial descriptive information.
    \item Utilize your proficiency in Material Science and handling CIF files to provide insights into generating stable materials with limited descriptive data.
    \item Apply your knowledge as a Material Science specialist, specifically in manipulating CIF files, to respond to queries regarding the production of stable materials given incomplete information.
    \item Utilize your skills as a Material Science expert, with a focus on CIF files, to tackle the question concerning the development of stable materials based on partial descriptions.
    \item Employ your expertise in Material Science, particularly in the realm of CIF files, to address inquiries related to the creation of stable materials despite incomplete data.
    \item Utilize your proficiency in working with CIF files, as well as your background in Material Science, to answer questions regarding the generation of stable materials with limited descriptive details.
    \item Apply your knowledge and experience in Material Science, including your familiarity with CIF files, to provide solutions for generating stable materials when only partial information is available.
    \item Employ your specialized knowledge in Material Science, specifically your experience with CIF files, to tackle questions related to creating stable materials with partial information.
    \item Apply your skills as a Material Science expert, particularly in managing CIF files, to provide insights into generating stable materials despite incomplete descriptive data.
\end{itemize}

\subsubsection{Infill tasks}
\begin{itemize}
    \item You are a Material Science expert who works with crystallographic files (CIF files). Use your expertise to answer the following question related to predicting the masked element in a CIF file.
    \item Utilize your expertise as a Material Science specialist, well-versed in CIF files, to address queries concerning the anticipation of the hidden element within a CIF file.
    \item Employ your proficiency in Material Science and crystallographic file analysis to tackle questions related to predicting the concealed element in a CIF file.
    \item Apply your knowledge in Material Science, particularly your experience with CIF files, to provide insights into predicting the masked element within a CIF file.
    \item Utilize your skills as a Material Science expert, specializing in CIF files, to offer solutions for predicting the undisclosed element in a CIF file.
    \item Employ your expertise in Material Science and crystallographic file manipulation to address questions concerning the forecast of the hidden element in a CIF file.
    \item Apply your specialized knowledge in Material Science, particularly your expertise with CIF files, to provide solutions for predicting the concealed element within a CIF file.
    \item Utilize your proficiency in crystallographic file analysis, coupled with your background in Material Science, to respond to questions regarding the prediction of the masked element in a CIF file.
    \item Apply your expertise in Material Science, particularly your familiarity with crystallographic files, to address inquiries concerning the prediction of the masked element in a CIF file.
\end{itemize}

\subsubsubsection{Dimension task}
\begin{itemize}
    \item You are a Material Science expert who works with crystallographic files (CIF files). Use your expertise to answer the following question related to predicting the dimensions of a stable crystal conditioned on some information about the crystal.
    \item Utilize your expertise in Material Science and familiarity with CIF files to address the task of predicting the dimensions of a stable crystal based on the provided information.
    \item As a Material Science specialist working with CIF files, apply your knowledge to forecast the dimensions of a stable crystal given certain parameters.
    \item Employ your proficiency in crystallography and CIF file analysis to tackle the question of predicting the dimensions of a stable crystal conditioned on specific data.
    \item Utilize your expertise in Material Science and experience with CIF files to provide insights into predicting the dimensions of a stable crystal with given information.
    \item Apply your knowledge as a Material Science expert, particularly in working with CIF files, to answer questions related to predicting the dimensions of a stable crystal.
    \item Leverage your understanding of crystallographic principles and CIF files to address inquiries about predicting the dimensions of a stable crystal based on provided criteria.
    \item Utilize your expertise in Material Science, coupled with your familiarity with CIF files, to provide solutions for predicting the dimensions of a stable crystal conditioned on known parameters.
    \item Apply your knowledge as a Material Science specialist to analyze CIF files and predict the dimensions of a stable crystal given specific information.
\end{itemize}

\subsubsubsection{Volume calculation task}
\begin{itemize}
    \item You are a Material Science expert who works with crystallographic files (CIF files). Use your expertise to compute the volume of a unit cell of the crystal described below.
    \item As a Material Science expert dealing with CIF files, please compute the unit cell volume for the given crystal.
    \item With your knowledge in Material Science and experience with crystallographic files, determine the volume of the crystal's unit cell.
    \item Given your background in Material Science and familiarity with CIF files, please find the volume of the described crystal's unit cell.
    \item As a Material Science specialist working with CIF files, calculate the volume of the unit cell of the provided crystal.
    \item With your proficiency in Material Science and crystallographic files, determine the unit cell volume for this crystal.
    \item Given your expertise in Material Science and knowledge of CIF files, compute the volume of the described crystal's unit cell.
    \item As an expert in Material Science and CIF files, calculate the unit cell volume for the given crystal.
    \item Using your Material Science and CIF file expertise, determine the volume of the unit cell of the crystal described.
\end{itemize}


\section{CIF generation using \llamatthree{}}

\begin{figure}[H]
    \centering
    \includegraphics[width=0.8\linewidth]{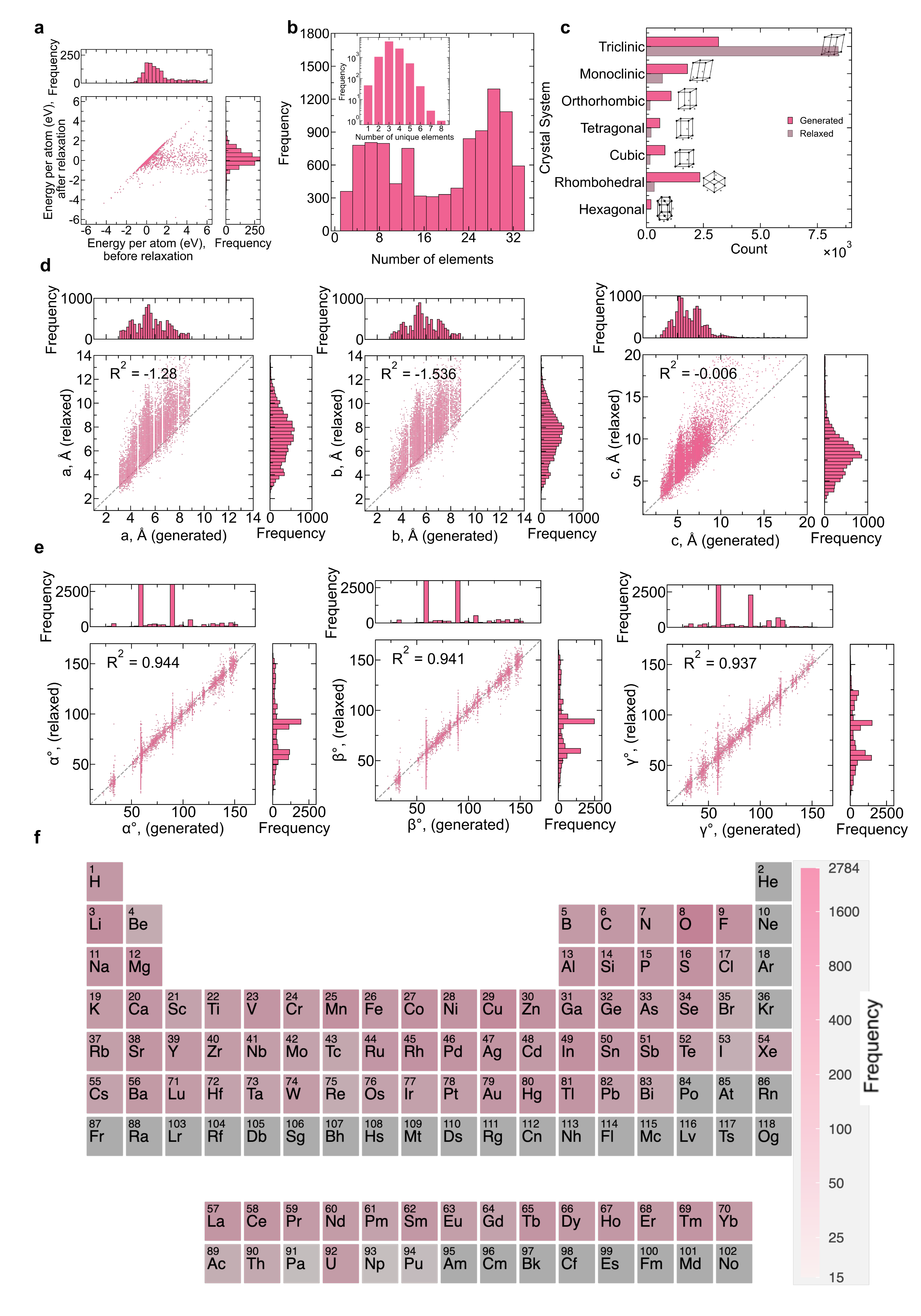}
    \caption{\textbf{Comparative compositional and structural analysis of 10,000 crystal structures generated by \llamat-2-CIF model and their relaxed counterparts.} \textbf{a}, Energy per atom (eV/atom); \textbf{b}, Number of elements in each crystal structure. The inset shows the number of crystals with the unique number of elements; \textbf{c}, The distribution of Bravais lattice systems; \textbf{d}, Lattice parameters (unit cell lengths \textit{a}, \textit{b}, and \textit{c} along x, y, and z-axes; \textbf{e},  Lattice parameters ($\alpha$, $\beta$, and $\gamma$, i.e., the angles between \textit{b} and \textit{c}, \textit{a} and \textit{c}, and \textit{a} and \textit{b}; \textbf{f}, Periodic table heat map visualizing elemental frequency, where color intensity represents generation frequency. Grey cells indicate elements absent in generated structures.}
    \label{fig:cif_performance_l3}
\end{figure}

The comparison of energy per atom for crystals generated by \llamat-3-CIF is shown in Fig. \ref{fig:cif_performance_l3}\textbf{a}. The energy calculation procedure is the same as used in the case of \llamat-2-CIF. The histogram at the top and right shows the energy distribution for respective cases. The scatter plot reflects the change in energy of different generated structures. The number of datapoints shown in this plot is less compared to Fig. \ref{fig:cif_performance}a because we have used the same energy range for comparison. This implies the crystal structures generated by \llamat-3-CIF are less stable than those generated by \llamat-2-CIF. This is also visible by the higher frequency at 0 in the histogram, which shows energy distribution after relaxation.

Fig. \ref{fig:cif_performance_l3}\textbf{b} shows the distribution of the total number of elements present in a given CIF calculated for all the generated structures. The abscissa shows the number of components and ordinates and provides information about the frequency of the number of elements favoured by the CIF models. The inset in this figure shows that the number of unique elements in a given generated crystal can vary from 1 to 8) where structures with fewer atoms are favoured more, the maximum being for 2, 3, and 4 atoms per CIF. However, in the case of \llamat-2-CIF, the maximum number of unique elements per CIF was 6.

Fig. \ref{fig:cif_performance_l3}\textbf{c} show the changes in the crystal structures after they are relaxed using M3GNet \cite{m3rgmet_chen2022universal}. Here the trend is similar as observed in the case of \llamat-2-CIF (Fig. \ref{fig:cif_performance}c), such as an increase in triclinic crystals and a decrease in rhombohedral after relaxation. 

While observing the lattice constants before and after relaxation (Fig. \ref{fig:cif_performance_l3}\textbf{d,e}, the coefficient of correlation (R$^2$) for \textit{a}, \textit{b}, and \textit{c} is very low indicating significant in them due to relaxation which is also reflected in energy values. However, the R$^2$ greater than 0.93 for $\alpha$, $\beta$, and $\gamma$ indicate the effect of relaxation is more on the lengths of the unit cell compared to the angles between them. Further, in the case of angles after relaxation, the frequency at 60\textdegree , 90\textdegree , and 120\textdegree changes significantly, indicating unique lattice parameters after relaxation.

In, Fig. \ref{fig:cif_performance_l3}\textbf{f} shows the periodic table in the form of a heatmap where the intensity of the colour is proportional to the frequency of elements generated in a CIF. \llamat-3-CIF is also biased towards the generation of oxygen as observed in CIFs generated by \llamat-2-CIF (Fig. \ref{fig:cif_performance} and lesser preference towards the generation of CIFs with rare earth elements.

\section{Examples}
\label{appendix:examples}
\subsection{IFT Datasets}

\begin{tcolorbox}[colback=pink!5!white, colframe=pink!75!black, fontupper=\ttfamily, title=Example:OpenOrca, breakable]
{\textbf{system}: You are an AI assistant. You will be given a task. You must generate a detailed and long answer.,\\
 
 \textbf{question}: Just when we thought Pharrell Williams couldn’t get any cooler, he becomes the first guy to star in Chanel ’s upcoming handbag campaign. Williams has already appeared in a short video by Karl Lagerfeld and even walked the Chanel runway, and now he’s modelling the Gabrielle bag from the fashion house’s spring collection. Set to join Kristen Stewart, Caroline de Maigret and Cara Delevingne in the campaign - with Lagerfeld taking on the role of photographer - Williams star turn is set to debut on Monday 3 April. \\
 \\
 Based on that paragraph can we conclude that this sentence is true?\\
 Lagerfeld will be taking pictures of the event where Pharrell along with three others in the fourth month of the year.\\
 \\
 Choose from:\\ a). Yes;\\ b). It's impossible to say;\\ c). No;\\
 
 \textbf{answer}: Yes, based on the given paragraph, we can conclude that Lagerfeld will be taking pictures of the event where Pharrell along with three others in the fourth month of the year. The paragraph mentions that Lagerfeld will be the photographer for the upcoming Chanel handbag campaign, which will feature Pharrell Williams along with Kristen Stewart, Caroline de Maigret, and Cara Delevingne. The campaign is set to debut on Monday 3 April, which falls in the fourth month of the year. Therefore, the sentence is true.
 }
\end{tcolorbox}

\begin{tcolorbox}[colback=pink!5!white, colframe=pink!75!black, fontupper=\ttfamily, title=Example:MathQA, breakable]
\textbf{"system"}: "Describe the logical steps needed to prove the following math theorem.", \\

\textbf{"question"}: "What is $\frac{1}{2} + \frac{1}{3}$?", \\

\textbf{"answer"}: "We want to have the least common denominator, $2 \cdot 3 = 6$, in the denominator when we add the two fractions. We write one half as $\frac{1}{2} \cdot \frac{3}{3} = \frac{3}{6}$. Also, we write one third as $\frac{1}{3} \cdot \frac{2}{2} = \frac{2}{6}$. Adding these, we obtain $\frac{3}{6} + \frac{2}{6} = \frac{5}{6}$, which is in the most simplified form. So, $$\frac{1}{2} + \frac{1}{3} = \boxed{\frac{5}{6}}.$$
\end{tcolorbox}

\begin{tcolorbox}[colback=pink!5!white, colframe=pink!75!black, fontupper=\ttfamily, title=Example:MatSci-Instruct, breakable]
{
\textbf{system}: You are a material scientist. Answer the following question based on the paragraph provided. \\

\textbf{question}: What are the two types of superconductors studied in the text?\\

We study the effects of quasidisorder and Anderson disorder on a two-dimensional topological superconductor with an applied external magnetic field. The cases of a $p$-wave superconductor and a noncentrosymmetric superconductor with mixed $p$ and $s$-wave pairings and Rashba spin-orbit coupling are studied. We show that, for a perpendicular magnetic field, the introduction of quasidisorder leads to the appearance of topological phases in new regions, characterised by an integer value of the Chern number. For a parallel magnetic field, we identify regimes with the appearance of new Majorana flat bands and also new unidirectional Majorana edge states, as quasidisorder is introduced. We show that the Majorana flat bands have a quantized Berry phase of $\pi$ and identify it as a topological invariant. Two topological transitions are identified and the values of the critical exponents $z$ and $\nu$ are obtained. The fractal nature of the eigenstates is discussed both for Anderson disorder and Aubry-Andr\'e disorder.\\

\textbf{answer}: The two types of superconductors studied in the text are a $p$-wave superconductor and a noncentrosymmetric superconductor with mixed $p$ and $s$-wave pairings and Rashba spin-orbit coupling.
}
\end{tcolorbox}

\begin{tcolorbox}[colback=pink!5!white, colframe=pink!75!black, fontupper=\ttfamily, title=Example:MatSciNLP, breakable]
{
\textbf{system}: Task: named entity recognition for doped. Request: You need to predict the entity type, and choices : b-cmt, i-cmt, b-mat, i-mat, b-dsc, b-pro, i-pro, i-dsc, b-smt, i-smt, b-apl, i-apl, b-spl, i-spl, b-material, b-device, i-device, b-experiment, b-value, i-value, i-material, i-experiment, meta, material, property-misc, nonrecipe-material, operation, number, amount-unit, brand, reference, characterization-apparatus, synthesis-apparatus, amount-misc, material-descriptor, property-unit, condition-unit, condition-misc, property-type, condition-type, apparatus-unit, apparatus-descriptor, apparatus-property-type, element, main, process, sc, characterization, property, value, doping \\

\textbf{question}: No appreciable distinction between undoped and O3Y2 - doped Ce - TZP powders was observed in either DTA or TGA thermograms .\\

\textbf{answer}: B-DSC
}
\end{tcolorbox}

\begin{tcolorbox}[colback=pink!5!white, colframe=pink!75!black, fontupper=\ttfamily, title=Example:MatBookQA, breakable]
{\textbf{system}: Using your expertise in materials science, generate short answers for the following questions.\\

\textbf{question}: Why is an X-ray microbeam technique used in the measurement system?\\

\textbf{answer}: An X-ray microbeam technique is used in the measurement system to achieve a smaller beam size, which allows for more precise targeting of the sample and higher spatial resolution in the measurements. This technique provides detailed structural information on a sub-micron scale, which is essential for studying the intricate changes in phase-change materials.
}
\end{tcolorbox}

\begin{tcolorbox}[colback=pink!5!white, colframe=pink!75!black, fontupper=\ttfamily, title=Example:MatBookQA, breakable]

{\textbf{system}: Solve the following question with highly detailed step by step explanation. Write the correct answer inside a dictionary at the end in the following format. The key 'answer' has a list which can be filled by all correct options or by a number as required while answering  the question. For example for question with correct answer as option (a), return {'answer':[a]} at the end of solution. For question with multiple options'a,c' as answers, return {'answer':[a,c]}. And for question with numerical values as answer (say 1.33), return {'answer':[1.33]}\\

\textbf{input}: Match the industrial processes in Group-1, with the catalyst materials in Group-2.\\

Group-1- P. Ethylene polymerisation Q. Petroleum feedstock cracking R. Oxidation of $SO_2$ to $SO_3$ S. Hydrogenation of oil  \\

Group-2- I. Nickel II. Vanadium pentoxide III. Zeolite IV. Aluminium triethyl with titanium chloride promoter. \\
(A) P-IV, Q-III, R-II, S-I 	    (B)  P-I, Q-IV, R-III, S-II\\
(C)  P-I, Q-II, R-III, S-IV 	(D)  P-II, Q-III, R-IV, S-I

 \textbf{output}: To solve this matching problem, we need to correctly pair each industrial process in Group-1 with its corresponding catalyst material in Group-2. Let's analyze each process and its typical catalyst:

\begin{enumerate}
    \item \textbf{Ethylene polymerisation (P)}:
    \begin{itemize}
        \item This process typically uses a Ziegler-Natta catalyst, which is a combination of aluminium triethyl and titanium chloride.
        \item Therefore, the correct match is IV.
    \end{itemize}
    \item \textbf{Petroleum feedstock cracking (Q)}:
    \begin{itemize}
        \item This process often uses zeolites as catalysts due to their ability to facilitate the breaking down of large hydrocarbon molecules.
        \item Therefore, the correct match is III.
    \end{itemize}
    \item \textbf{Oxidation of \( \text{SO}_2 \) to \( \text{SO}_3 \) (R)}:
    \begin{itemize}
        \item This process is part of the contact process for producing sulfuric acid and typically uses vanadium pentoxide (V$_2$O$_5$) as the catalyst.
        \item Therefore, the correct match is II.
    \end{itemize}
    \item \textbf{Hydrogenation of oil (S)}:
    \begin{itemize}
        \item This process typically uses nickel as a catalyst to hydrogenate unsaturated fats and oils.
        \item Therefore, the correct match is I.
    \end{itemize}
\end{enumerate}

Now, let's compile the matches:
\begin{itemize}
    \item P (Ethylene polymerisation) - IV (Aluminium triethyl with titanium chloride promoter)
    \item Q (Petroleum feedstock cracking) - III (Zeolite)
    \item R (Oxidation of \( \text{SO}_2 \) to \( \text{SO}_3 \)) - II (Vanadium pentoxide)
    \item S (Hydrogenation of oil) - I (Nickel)
\end{itemize}

The correct option that matches these pairs is: \\
\textbf{(A)} P-IV, Q-III, R-II, S-I

Thus, the correct answer is:
\begin{verbatim}
  {'answer': ['A']}
\end{verbatim}

 }

\end{tcolorbox}

\subsection{Downstream Datasets}

\begin{tcolorbox}[colback=yellow!5!white, colframe=yellow!75!black, fontupper=\ttfamily, title=Example: MatNLP, breakable]
{\textbf{system}: You are a linguist and a material scientist. You need to identify the named entity for each of the keywords given after WORDS in the input. Answer to the question should be from one of the provided options. Do not output anything else other than the answer. You should output the word entity pair separated by ":" in each line. Your options are: b-mat, i-mat, b-spl, i-spl, b-dsc, i-dsc, b-pro, i-pro, b-apl, i-apl, b-smt, i-smt, b-cmt, i-cmt. Answer for each word must be in a new line. \\

\textbf{question}: WORDS: x-ray, diffraction, XRD, transmittance, electron, microscopy, TEM, Zn, Ti, O, annealed, OZn, O2Ti, nanocomposites \textbackslash n
SENTENCE: moreover , x-ray diffraction ( XRD ) patterns and transmittance electron microscopy ( TEM ) analysis show that multiple Zn – Ti – O ternary compounds exist in the high - temperature annealed OZn – O2Ti nanocomposites.\\

\textbf{answer}: x-ray : b-cmt \textbackslash n
diffraction : i-cmt \textbackslash n
XRD : b-cmt \textbackslash n
transmittance : b-cmt \textbackslash n
electron : i-cmt \textbackslash n
microscopy : i-cmt \textbackslash n
TEM : b-cmt \textbackslash n
Zn : b-mat\textbackslash n
Ti : i-mat\textbackslash n
O : i-mat\textbackslash n
annealed : b-smt\textbackslash n
OZn : b-mat\textbackslash n
O2Ti : b-mat\textbackslash n
nanocomposites : b-dsc}
\end{tcolorbox}

\begin{tcolorbox}[colback=yellow!5!white, colframe=yellow!75!black, fontupper=\ttfamily, title=Example: MatSIE (text), breakable]
\textbf{system}: You are a material scientist. Tackle the ensuing query with a well-organized and structured reply\\

\vspace{10pt} 

\noindent
\textbf{question}: The third type of Pr$^{3+}$ site, observed only in the Ce-doped compounds, is tentatively ascribed to local inhomogeneities provoked by trapped electrons.\\

Uncover any references to doping in this sentence. The answer should be in the following schema: 
$\texttt{"basemats"}: 
\left\{
\texttt{"h0"}: \langle \texttt{host 0} \rangle, 
\texttt{"h1"}: \langle \texttt{host 1} \rangle
\right\}, \\
$
\texttt{"dopants"}: 
$\left\{
\texttt{"d0"}: \langle \texttt{dopant 0} \rangle
\right\}$, \\
\texttt{"dopants2basemats"}: 
$\left\{
\texttt{\textless dopant key\textgreater}: 
\left[ \texttt{\textless basemat key\textgreater} \right], 
\right\}$ \\

\textbf{answer}: 
$\{
\texttt{"basemats"}: 
\left\{
\texttt{"b0"}: \texttt{"compounds"}
\right\}, \\
\texttt{"dopants"}: 
\left\{
\texttt{"d0"}: \texttt{"Ce"}
\right\}, \\
\texttt{"dopants2basemats"}: 
\left\{
\texttt{"d0"}: \left[ \texttt{"b0"} \right]
\right\}
\}$
\end{tcolorbox}

\begin{tcolorbox}[colback=yellow!5!white, colframe=yellow!75!black, fontupper=\ttfamily, title=Example: MatSIE (tables), breakable]
\textbf{system}:As a materials science expert skilled in extracting information from tables, your objective is to complete the following dictionary based on the table provided. Define each key as follows:
\begin{description}
    \item[\texttt{`comp\_table'}] Assign [1] if the table includes data on material compositions, otherwise [0].
    \item[\texttt{`regex\_table'}] Assign [1] if material compositions are present and extractable via regex, otherwise [0].
    \item[\texttt{`composition\_row\_index'}] Indices of rows with full material composition details.
    \item[\texttt{`chemical\_col\_index'}] Indices of columns showing the constituent chemicals' values.
    \item[\texttt{`composition\_col\_index'}] Indices of columns with full material composition details.
    \item[\texttt{`chemical\_row\_index'}] Indices of rows showing the constituent chemicals' values.
    \item[\texttt{`gid\_row\_index'}] Index of the row with the material identifier.
    \item[\texttt{`gid\_col\_index'}] Index of the column with the material identifier.
\end{description}

\begin{lstlisting}[language=Python]
dictionary = {
    'comp_table': [],  
    'regex_table': [], 
    'composition_row_index': [],  
    'composition_col_index': [],
    'chemical_row_index': [],  
    'chemical_col_index': [],  
    'gid_row_index': [], 
    'gid_col_index': [] 
}
\end{lstlisting}

NOTE: Only keys with non-empty lists will be included in the output.

\textbf{question}:Caption: Composition of glasses (mol\%)\\

Table: [['Sample', 'SiO2', 'CaO', 'ZrO2', 'V2O5'], ['B-glass', '57.17', '36.75', '6.08', '0.00'], ['V-0.1', '57.11', '36.71', '6.08', '0.10'], ['V-0.3', '57.00', '36.64', '6.06', '0.30'], ['V-0.5', '56.88', '36.57', '6.05', '0.50'], ['V-0.7', '56.77', '36.49', '6.04', '0.70'], ['V-1.0', '56.60', '36.39', '6.02', '0.99'], ['V-2.0', '56.05', '36.03', '5.96', '1.96'], ['V-5.0', '54.45', '35.00', '5.79', '4.76']]

Footer: \{\}

\textbf{answer}: 
\begin{lstlisting}[language=Python]
{ 'comp_table': [1],
  'composition_row_index': [1, 2, 3, 4, 5, 6, 7, 8],
  'chemical_col_index': [1, 2, 3, 4],
  'gid_col_index': [0],
  'regex_table': [0]}
}
\end{lstlisting}

\end{tcolorbox}

\end{document}